\documentclass[aps,prb]{revtex4} 

\usepackage{psfig}

\input{macros.phys}

\def\sinf{\sigma_{\infty}}
\def\sinfbar{\bar{\sigma}_{\infty}}

\begin{document}

\title{Random tilings of high symmetry: II. Boundary conditions and numerical
studies}

\author{M. Widom}
\affiliation{Department of Physics, Carnegie Mellon University, 
Pittsburgh, PA  15213, USA.}

\author{N. Destainville}
\affiliation{Laboratoire de Physique Th\'eorique, UMR CNRS-UPS 5152,
Universit\'e Paul Sabatier,
31062 Toulouse Cedex 04, France.}

\author{R. Mosseri} \affiliation{Laboratoire de Physique Th\'eorique des
Liquides, Tour 24, Bo\^{\i}te 121, 4, Place Jussieu, 75252 Paris Cedex 05,
France.}
 
\author{F. Bailly}
\affiliation{LPSB-CNRS, 92195 Meudon Cedex, France.}

\date{\today}

\begin{abstract}
We perform numerical studies including Monte Carlo simulations of high
rotational symmetry random tilings.  For computational convenience,
our tilings obey fixed boundary conditions in regular polygons.  Such
tilings are put in correspondence with algorithms for sorting lists in
computer science. We obtain statistics on path counting and vertex
coordination which compare well with predictions of mean-field theory
and allow estimation of the configurational entropy, which tends to
the value 0.568 per vertex in the limit of continuous
symmetry. Tilings with phason strain appear to share the same entropy
as unstrained tilings, as predicted by mean-field theory. We consider
the thermodynamic limit and argue that the limiting fixed boundary
entropy equals the limiting free boundary entropy, although these
differ for finite rotational symmetry.
\end{abstract}

\maketitle

\section{Introduction}
\label{intro}
\setcounter{equation}{0}

A tiling is a filling, without gaps or overlaps, of a given region of
a $d$-dimensional Euclidean space, with tiles which differ according
to their shapes, sizes, and orientations. In the present paper, the
tiles are $d$-dimensional rhombohedra, which we will generically call
``rhombi'' in the following.  Each tile is the projection of a
$d$-dimensional face of a $D$ dimensional hypercube ($D>d$) into
$d$-dimensional space. The difference $D-d$ is known as the
codimension of the tiling, and we say that we are dealing with $D \ra
d$ tilings. Details of this construction can be found in a previous
paper~\cite{paperI}, hereafter referred to as ``paper I''. We mainly
focus on the case $d=2$.  A random rhombus tiling can be viewed as a
fluctuating membrane in this higher dimensional space~\cite{Elser},
the membrane being the union of hypercube faces (see paper
I~\cite{paperI}, section II.1).

Rhombus tilings are dual to de Bruijn grids~\cite{paperI,debruijn}. In
two dimensions, these grid lines pass through the midpoints of
parallel rhombus edges. Every rhombus edge orientation defines a
family of effectively parallel de Bruijn grid lines. De Bruijn grid
lines within a family never cross. In contrast, lines of different
families {\em do} cross, and their crossing defines a rhombus of the
tiling. There are $D$ rhombus edge orientations and hence $D$ families
of de Bruijn grid lines.

\begin{figure}[htb]
\begin{center}
\ \psfig{figure=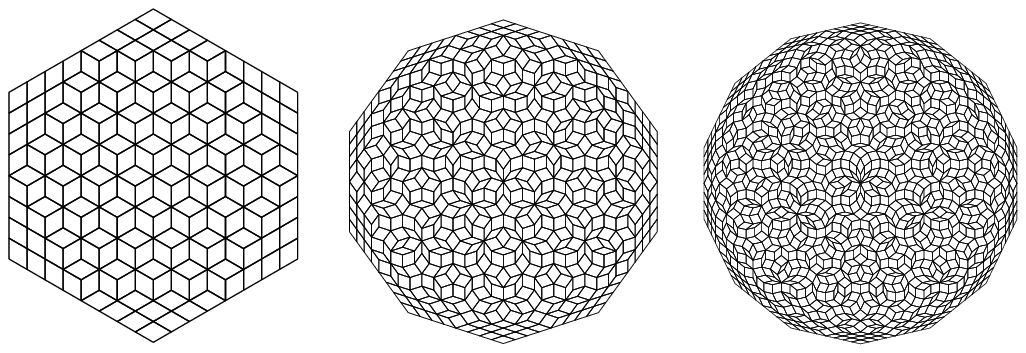,width=12.5cm} \
\end{center}
\caption{Fixed boundary $D \ra 2$ tilings with $D=3,5$ and $7$. The
  side lengths of the polygons are all $p=8$.}
\label{pavagesD2}
\end{figure}

Rhombus tilings provide simple models for
quasicrystals~\cite{Shechtman,Levine}, metal alloys that exhibit
rotational symmetry forbidden by conventional crystallography. One of
the key properties of random tilings~\cite{Henley91} is their
configurational entropy that may play a role in stabilizing the
quasicrystal state. The source of entropy is localized tile
rearrangements known as phason flips. Groups of three adjacent rhombi
may be permuted so that their perimeter remains fixed while their
shared vertex moves to a nearby point. Our present focus, as in paper
I, is the limit of high rotational symmetry obtained as $D \ra \infty$
at fixed $d$.  We examine this limit because of its intrinsic interest
and because it deepens our understanding of finite $D$ tiling models
relevant for real alloy systems.

As usual in statistical mechanics, tiling systems can have different
boundary conditions, such as free, fixed or periodic. For example, the
tilings in figures~\ref{pavagesD2}, \ref{ex40.2} or
\ref{DeBruijn4.2}, have fixed polygonal boundaries. It is believed that
free  and periodic boundary tilings reach equal entropies at the
thermodynamic limit.  Fixed boundary tilings, in contrast, exhibit
entropies that are strictly smaller than the free boundary
entropy. For example, the fixed hexagonal-boundary entropy of $3 \ra
2$ tilings (see Fig.~\ref{pavagesD2}, left) equals
0.261~(\cite{elsershape}), while the free boundary entropy equals
0.323~(\cite{Wannier}). This phenomenon can be
understood~\cite{elsershape,Grensing,matheux,Bibi97B} by inspection of
Fig.~\ref{pavagesD2}. The local entropy density displays a gradient
between crystalline regions near the boundary, where the entropy
density vanishes, and the central region, where the entropy density
reaches a maximum value equal to the free boundary quasicrystalline
value. Only at the very center of the tiling does the influence of the
boundary disappear.

Paper I~\cite{paperI} developed a mean-field theory for free boundary
tilings, applicable to the limit of high rotational symmetry.  Two
earlier papers~\cite{Widom97,Bibi00} presented initial studies of this
problem. The first paper~\cite{Widom97} proposed an upper bond on the
entropy in the limit of large $D$ and discussed problems associated
with the thermodynamic limit of tiling models. The second
one~\cite{Bibi00} presented a preliminary mean-field approach of the
entropy calculation. In the present paper we focus on fixed boundary
tilings because they are easy to simulate numerically. We present
results of our simulations, including an accurate estimate of the
limiting entropy density. We investigate the role of fixed boundary
conditions and argue for a boundary-condition-independent
thermodynamic limit in the limit $D \ra \infty$. Similarly, we show
that phason strain does not influence the entropy for large $D$.

The organization of this paper is as follows.  We start in
section~\ref{fixed} with a discussion of fixed boundary tilings, and
describe their relationship to interesting problems of pure
mathematics. Next, we review the problem of the thermodynamic limit in
section~\ref{thermo} where we argue that free and fixed boundary
tilings attain the same thermodynamic limit as $D \ra \infty$.  We
also explore finite $D$ corrections to the fixed boundary entropy and
their relation to the inhomogeneity of the tilings.  Then,
section~\ref{simul} describes our Monte Carlo simulations. In that
section we explore the entropy, path counting statistics, vertex
coordination statistics and the role of phason strain. We also confirm
numerically the results of section~\ref{thermo}. The paper is written
so that the reader does not need the notations of paper I, except in
the appendices.

\section{Fixed boundary tilings}
\label{fixed}
\setcounter{equation}{0}

Fixed boundary tilings such as those of figure~\ref{pavagesD2} possess
the most natural boundary conditions for tilings coded by
combinatorial objects such as generalized partitions or sorting
algorithms.  Even though these boundary conditions are unnatural for
bulk quasicrystals, this inconvenience is counter-balanced by the
powerful tools provided by such codings in terms of understanding of
tiling set structures and enumeration of tilings.  Furthermore, we
will demonstrate in this paper that fixed  and free boundary tilings
become equivalent in the large codimension limit, which will
justify {\em a posteriori} the use of fixed boundary conditions.

As already described in the introduction, such fixed boundary tilings
lack a proper thermodynamic limit for finite $D$ because the boundary
has a strong macroscopic effect on the whole
tiling~\cite{elsershape,Grensing,matheux,Bibi97B,4to3}. This leads to
a spectacular effect known as the ``arctic circle
phenomenon~\cite{matheux}'' in hexagonal ($D=3$) tilings, where the
tiling is periodic (and ``frozen'') outside a perfect circle at the
large size limit and random inside this circle. More generally and for
larger $D$, fixed boundary tilings have an effective codimension that
is smaller near the boundary than in the bulk. Very near the boundary
the effective codimension vanishes. The tiling becomes a crystalline
domain comprising a single tile type. The local entropy density
vanishes because no phason flips are possible. In the membrane
picture, this region corresponds to a flat area with large
tilt. Further from the boundary the effective codimension grows. The
variable codimension results in an entropy density gradient, growing
from zero at the boundary to the maximal, free boundary, entropy
density at the tiling center. Consequently, fixed boundary tilings
have a smaller total entropy per tile than free boundary tilings.

\begin{figure}[ht]
\begin{center}
\ \psfig{figure=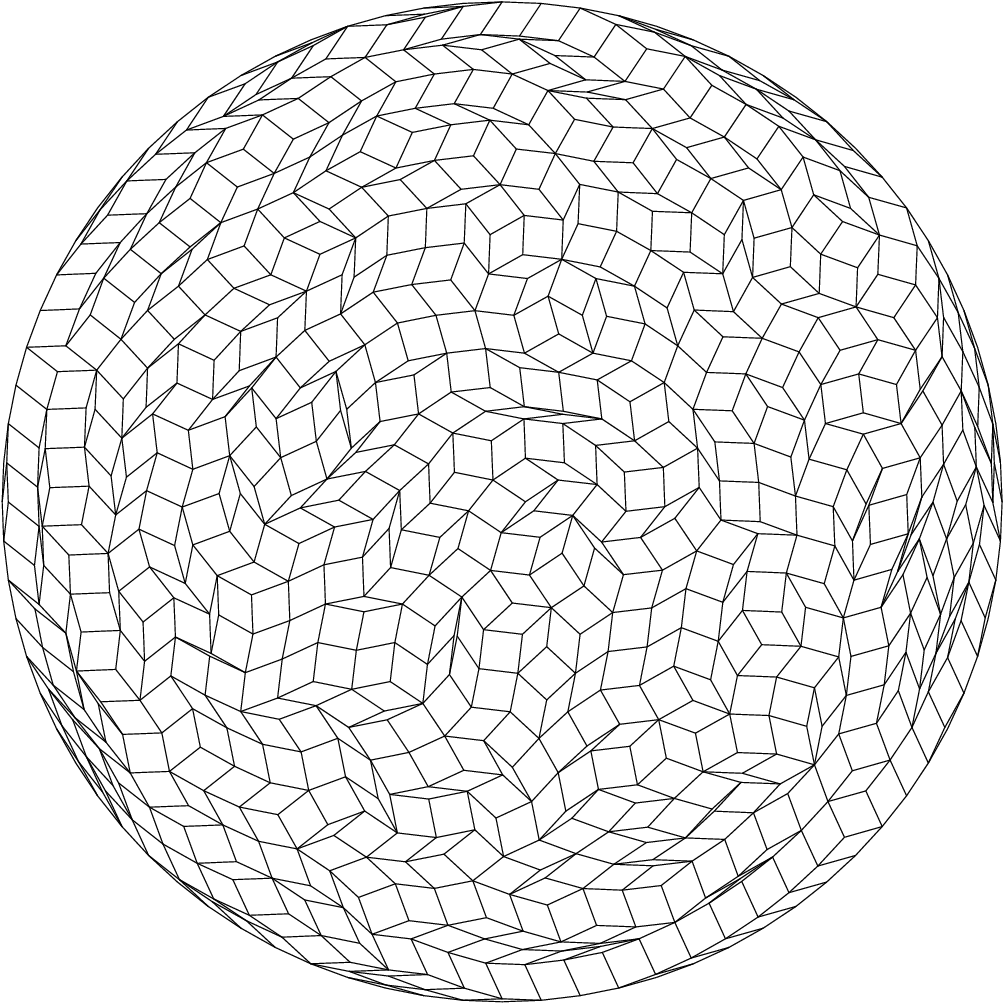,width=5cm} \
\end{center}
\caption{A $40 \ra 2$ fixed boundary tiling with one de Bruijn line in
  each family.}
\label{ex40.2}
\end{figure}

In contrast, we will show in section~\ref{thermo} that when $D$
becomes large (figure~\ref{ex40.2}), this heterogeneity diminishes and
a thermodynamic limit is restored. Since fixed boundary tilings are
easier to specify and manipulate for arbitrary $D$, they are the
appropriate tool to tackle large $D$ entropies or other statistical
properties.  In the following, the fixed boundary entropies per tile
will be denoted by $\bar{\sigma}_D$ and their limiting value by
$\sinfbar$. Corresponding values without the over-bar refer to free
boundary tilings.

In terms of de Bruijn dualization, any two lines of two different
families intersect inside a fixed boundary tiling~\cite{octo01} (the grid
is said to be ``complete''). Duality associates tile with
the intersection of two lines. Therefore if there are $k_i$ lines in
each family $F_i$, the number of tiles is
\begin{equation}
N_T = \sum_{1\leq i<j \leq D} k_i k_j.
\label{NT}
\end{equation}
If all $k_i$ are set to 1, $N_T = D(D-1)/2$.

\begin{figure}[ht]
\begin{center}
\ \psfig{figure=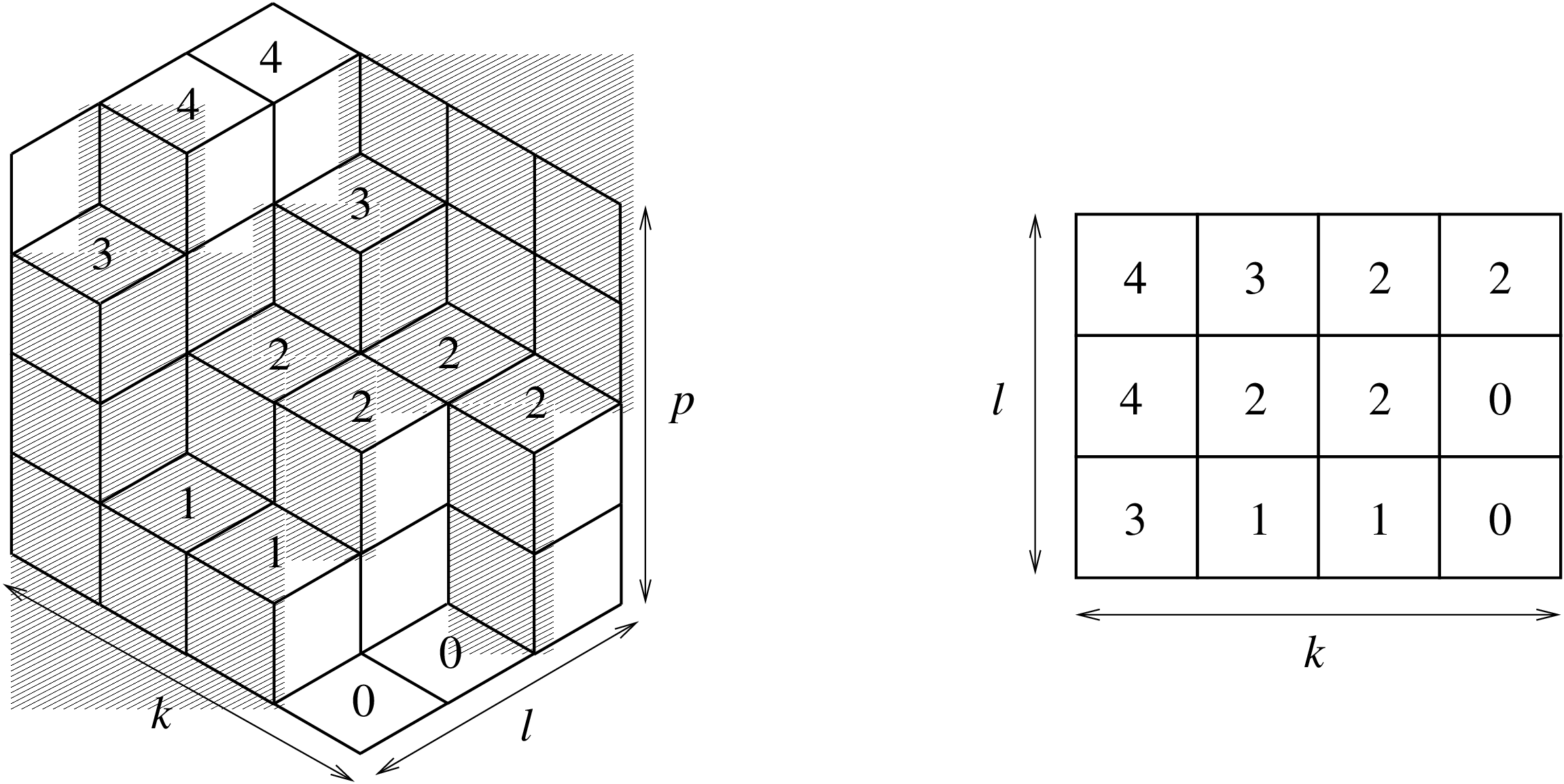,height=4.5cm} \
\end{center}
\caption{Three-dimensional representation of an hexagonal tiling filling
a centrally symmetric hexagon (left). This object can be seen
as a staking of cubes. The height of the different stacks can be
arranged in an array (right). They are decreasing in rows and columns. This
array together with these order relations constitutes a plane
partition problem.}
\label{ex.part}
\end{figure}

The relationship between tilings and partitions has been widely
explored in references~\cite{elsershape,octo01,Mosseri93B,Bibi97}. The
idea is to code a random rhombus tiling by an array of integers
satisfying certain ordering constraints.  Figure~\ref{ex.part}
illustrates this point in the simple hexagonal case~\cite{elsershape}:
there is a one-to-one correspondence between hexagonal tilings filling
a centrally symmetric hexagon of sides $k,l$ and $p$ on the one hand
and sets of integers arranged in a rectangular array $k \times l$,
decreasing in each row and each column, on the other hand. These
integers are bounded between zero and the side length $p$. They
decrease monotonically. Such a set of integers is called a plane
partition.

This point of view can be extended to any $D \ra d$
problem~\cite{octo01,Mosseri93B,Bibi97}: any $D \ra d$ tiling can be
coded with a generalized partition on an array related to a $D-1 \ra
d$ tiling. This correspondence is also one-to-one. However, tilings
coded by partitions have fixed polytopal boundary conditions. In two
dimensions, the boundaries are centrally symmetric $2D$-gons (see figure~\ref{pavagesD2}).

We now present an analogy between fixed boundary tilings and
algorithms for sorting lists~\cite{octo01}. Although Computer Science is
not our motivation, we are interested in the same enumeration problem
\cite{Knuth92,Bjorner93}. More generally, the rich topological and
combinatorial properties of random tilings make them an active field
of research in pure mathematics, combinatorics and computer
science\cite{matheux,Kenyon93,Elnitsky97,Bailey97}. We shall use results from
these fields throughout this paper.

In the sorting language, a {\em comparator} $[i;j]$ acts on a list
$(x_1, x_2,\ldots, x_D)$ of numbers as follows: $x_i$ and $x_j$ are
respectively replaced by $\min(x_i,x_j)$ and $\max(x_i,x_j)$.
Following Knuth \cite{Knuth92}, we call a {\em complete} sorting
algorithm a sequence of such comparators which sorts in the increasing
order any list of real numbers $(x_1,x_2,\ldots,x_D)$. This sorting
algorithm will be called {\em primitive} if each comparator can be
written $[i;i+1]$. We also suppose that this algorithm is not
redundant, that is to say it does not contain any comparator $[i;j]$
that could be suppressed because previous comparators already insure
that $x_i \leq x_j$. Knuth shows that a sequence of comparators is a
sorting algorithm if it correctly sorts the completely reversed list
$(D,D-1,\ldots,1)$. This means that a complete primitive sorting
algorithm is a sequence of comparators $[i;i+1]$ that transforms the
list $(D,D-1,\ldots,1)$ into the list $(1,2,\ldots,D)$.

\begin{figure}[ht]
\begin{center}
\ \psfig{figure=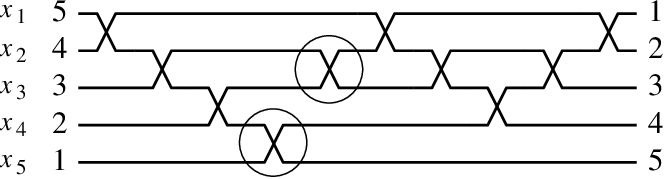,height=2.5cm} \
\end{center}
\caption{A diagram associated with a sorting algorithm acting on five
element lists. A line follows a number during the sorting process.
Each pair of lines cross, only once. Each crossing represents a
comparator $[i;i+1]$.}
\label{sort}
\end{figure}

Such an algorithm has a diagrammatical representation in which the $D$
variables $x_i$ are represented by $D$ horizontal lines. Each
comparator $[i;i+1]$ is represented by a crossing 
between lines $i$ and
$i+1$. Figure~\ref{sort} illustrates this construction. A continuous
line follows a number during the sorting process.  Since every number
must be compared to every other number, and since there is not any
redundancy, every line crosses every other line, only once. There are
$\simp{D}{2}$ crossings.

We now establish the link between sorting algorithms and the de Bruijn
representations~\cite{paperI,debruijn} of $D \ra 2$ tilings. Each line
of the diagram~\ref{sort} represents a de Bruijn line and crosses
every other line exactly once. However, different sorting algorithms
sometimes represent the same de Bruijn grid, since only the crossing
topology is meaningful. For example, in figure~\ref{sort}, the fourth
and the fifth comparator ({\em i.e.}  [4;5] and [2;3]) are applied in
this order (these comparator are circled in the figure). If they were
applied in the reverse order, the algorithm would be different whereas
the de Bruijn grid would be the same.

Therefore we define equivalence classes of sorting algorithms
\cite{Knuth92,Elnitsky97}. We say that two successive comparators
$[i;i+1]$ and $[j;j+1]$ commute if $| i-j | >1$. Two algorithms are
equivalent if they differ by a finite number of comparator
commutations. Equivalence classes of $D$-element sorting algorithms
are in one-to-one correspondence with $D$-family grids with one line
per family, and therefore with tilings inscribed in polygons of side
1.  Following Knuth, we denote by $B_D$ this number of equivalence
classes, whereas the number of algorithms is denoted by $A_D$ ($A_D
\geq B_D$). Since each pattern of crossings (equivalence class of
algorithms) defines a tiling, the tiling entropy density per tile is
\begin{equation}
\label{sigmaD}
\bar{\sigma}_D={{1}\over{N_T}} \log{B_D}.
\end{equation}

\begin{figure}[ht]
\begin{center}
\ \psfig{figure=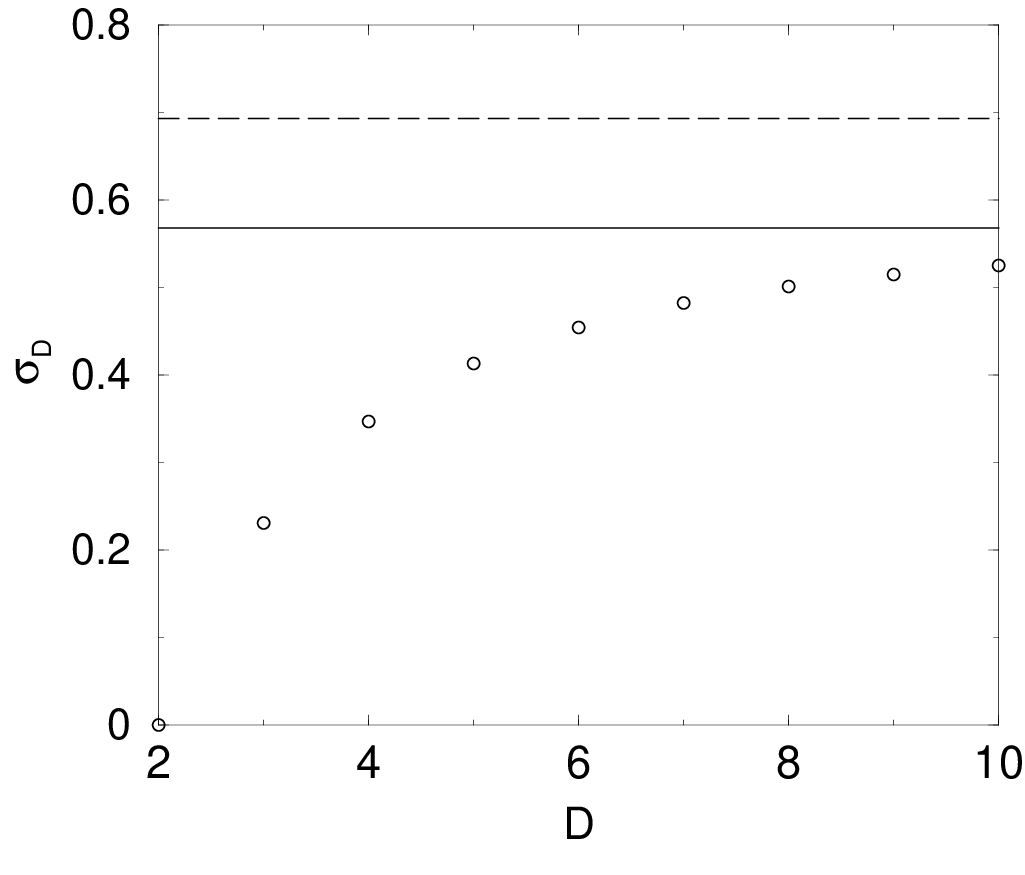,width=7cm} \
\end{center}
\vspace{-4mm}
\caption{Entropy per tile $\bar{\sigma}_{D}$ of fixed boundary
tilings ($p=1$). We indicate the limiting values $0.568$ (full line)
and the close upper bound $\log 2 \simeq 0.693$ (dashed line).}
\label{plot.pavages.finis}
\end{figure}

Sorting algorithms are easy to manipulate and provide efficient
enumeration numerical tools.  This analogy provides us with the number
$B_D$ of $D \ra 2$ tilings for small values of $D \le 10$ from the
work of Knuth \cite{Knuth92} as extended by us~\cite{Widom97} (see
Table~\ref{pavages.finis}). Unfortunately it is not possible to reach
the next value ($D=11$) for the foreseeable future using our current
algorithms.  Figure~\ref{plot.pavages.finis} suggests that the entropy
$\sinfbar(p=1)$ has a well defined limit when $D$ tends to infinity.
Indeed, Knuth~\cite{Knuth92} gives lower and upper bounds on $B_D$
from which we get
\begin{equation}
{1 \over 3} \log 2 \leq \lim_{D \ra \infty} \bar{\sigma}_{D}
\leq 2 \log 2. 
\end{equation}
Moreover, Bj\"orner (see reference \cite{Bjorner93}, p.~270) 
derives a better upper bound: 
\begin{equation}
\lim_{D \ra \infty}  \bar{\sigma}_{D} \leq 1.44 \log 2.
\end{equation}
In the following section, we argue that large $D$, fixed
boundary tilings, have the same entropy per tile as free boundary
ones. Since we demonstrated in paper I~\cite{paperI} that for
free boundary entropies $\lim_{D \ra \infty} \sigma_D \leq
\log{2}$, we finally get a better bound:
\begin{equation}
\label{Ubound}
\lim_{D \ra \infty} \bar\sigma_D  \leq \log{2}.
\end{equation}

This point of view generalizes to systems with more than one line per
de Bruijn family~\cite{octo01}, leading to the definition of {\em
partial} sorting algorithms for merging pre-ordered lists of
numbers. Suppose that we have $D$ families of $k_i$ numbers each
($i=1,\ldots,D$), and that in each family the numbers are presorted in
increasing order. Algorithms which order the union of these sets of
numbers are called partial sorting algorithms. The ideas are
essentially the same as in the previous case, except that, since the
numbers of a given family are already ordered, the corresponding lines
do not cross within the family. The corresponding diagram is similar
to a de Bruijn grid with $D$ families of lines, $k_i$ lines in each
family. The tilings are equivalence classes of such algorithms.  They
are inscribed in polygons of sides $k_1,\ldots,k_D$. In
reference~(\cite{octo01}), we derive an analytic expression for
$A_D(k_1,\ldots,k_D)$. However, it proves impossible to calculate
$B_D$ analytically.

The diagonal case, with $k_1=k_2=\cdots=k_D=p$ will be of special interest
below. Let $B_D(p)$ denote the number of such tilings.  Using the
sorting algorithm analogy, we computed the numbers of tilings
inscribed in polygons of side 2 ($B_D(2)$) for small values of $D$
(see Table~\ref{pavages.finis2}).

\section{Thermodynamic limit}
\label{thermo}
\setcounter{equation}{0}

In this section, we argue that the fixed boundary entropy in the large
$D$ limit equals the free boundary entropy and is independent of the
phason strain. Proofs are provided in appendices
\ref{fixe} and \ref{encadre}. Free boundary and fixed boundary entropies can be
compared when both are known. For hexagonal tilings with $D=3$, the
free and fixed entropies are respectively 0.323~\cite{Wannier} and
0.261~\cite{elsershape}. For octagonal tilings with $D=4$ the values
are 0.434~\cite{lpw} and 0.36(1)~\cite{octo01}. Therefore the relative
difference $(\sigma-\bar{\sigma})/\sigma$ decreases between $D=3$ and
$D=4$, consistent with a vanishing difference at large $D$.
Exact data on $p=1$ and $p=2$ tilings
(Tables~\ref{pavages.finis} and~\ref{pavages.finis2}) suggest that
when $D$ becomes large, the entropy becomes independent of $p$.

A glance at figure~\ref{pavagesD2} suggests that when $D$ becomes
large, the faceted regions of fixed boundary tilings, which are caused
by the strong influence of the boundary \cite{Bibi97}, occupy a
vanishing fraction of the tiling area close to the boundary. In these
regions, few families of de Bruijn lines cross, so only a small number
of tile types are present. The central region of the tiling, where a
large number of de Bruijn line families cross, becomes dominant as $D$
increases. The coarse-grained entropy density in the central region
approaches the large $D$ free boundary entropy $\sinf$.

More precisely, let $(x_i)$ be any set of numbers $0 < \alpha
\leq x_i \leq \beta$, where $\alpha$ and $\beta$ are fixed positive
real numbers. Consider fixed boundary tilings of a $2D$-gon with edge
lengths $k_i=x_i p$ as $p \ra \infty$. We impose such bounds on the
side lengths in order to be sure that our tilings are effectively
large codimension ones (see also paper I~\cite{paperI} (section II.3)).
The tile fractions are
\begin{equation}
\bar{n}_{ij} = \frac{k_i k_j}{N_T} = \frac{k_i k_j}{k_1 k_2 + \ldots +
k_{D-1} k_D},
\end{equation}
owing to eq.~(\ref{NT}). In an unstrained tiling, where all $k_i$ are
equal,
\begin{equation}
\bar{n}_{ij}^* = \frac{2}{D(D-1)}.
\end{equation}
These quantities differ noticeably from their free boundary 
counterparts~\cite{paperI}.
The condition $0 < \alpha \leq x_i \leq \beta$ imposes
\begin{equation}
\bar{n}_{ij} \geq \frac{\alpha^2}{\beta^2} \; \bar{n}_{ij}^*
= \frac{\alpha^2}{\beta^2} \; \frac{2}{D(D-1)}.
\end{equation}
This lower bound ensures that no tile fraction is vanishingly small and
therefore that the tiling is really a large codimension one rather
than a small codimension one with few defect lines added in.

Under this condition, appendix~\ref{fixe} proves that the
sequence $\bar{\sigma}_D$ reaches a thermodynamic limit
$\bar{\sigma}_{\infty}$ that does not depend on the set
$(x_i)$. Furthermore, these tilings are locally equivalent to free
boundary high symmetry tilings. The local entropy density thus
approaches $\sinf$, and $\bar{\sigma}_{\infty} = \lim_{D \ra \infty}
\bar{\sigma}_D = \sinf$. Moreover, this result remains valid if $D$
becomes large while $p$ is held fixed (even in the extreme case
$k_i=p=1$).  In particular, the limits $D \ra \infty$ and $p \ra
\infty$ commute with each other. The proofs of all these results are
given in appendix~\ref{fixe}.  This insensitivity of entropy to
boundaries could be anticipated since we know that phason elastic
constants vanish at large $D$~[\cite{paperI}] and that, because of the
regular character of large $D$ tilings established in
appendix~\ref{encadre}, the phason gradient $\vect{E}$ is nearly
everywhere bounded.

We find in appendix~\ref{fixe} that large $D$ tilings have a central
region holding half of the tiles, which contains lines of all de
Bruijn families and all tile species and which is nearly homogeneous
and strain-free.  Away from this central region, the tiling no longer
contains lines of all de Bruijn families. We define the effective
(coarse-grained) dimension $D_{eff}$ as the number of de Bruijn
families present in a small tiling patch, and the effective
codimension $c_{eff}=D_{eff}-2$. This effective codimension decreases
from $D-2$ at the center to 0 at the boundary.
For diagonal tilings ($x_i=1$ so that $k_i=p$ for all $i$), where the
tiling has a circular symmetry at the large $D$ limit, the effective
codimension $c_{eff}$ varies as
\begin{equation}
c_{eff}(r) = (D-2) \gamma (r/R) 
\end{equation}
where $r$ is the distance to the center and $R$ is the radius of the
tiling.  The function $\gamma(r/R)$ varies from the value
$\gamma(0)=1$ at the tiling center to $\gamma(1)=0$ at the boundary.
For large $D$, $c_{eff}(r)$ diverges for all $r<R$, so the local
entropy density approaches $\sigma_{\infty}$ for all $r<R$.

This dependence of $c_{eff}$ on the location in the tiling can be used
to estimate finite size corrections to $\bar{\sigma}_D$.  Consider a
fixed value of $D\gg 2$, and find the number of tiles in the regions
where $c_{eff} \leq c_0$, with $c_0\ll D-2$ some fixed value. These outer
regions form an annulus $R_{an} \leq r \leq R$. Inside the radius
$R_{an}$, the mean entropy density is nearly $\sigma_{\infty}$,
but in the annulus the mean entropy density is
$\sigma_{an}<\bar{\sigma}_D$.  We can estimate the mean entropy
density on the whole tiling of radius $R$ as
\begin{equation}
\bar{\sigma}_D \simeq n \sigma_{an} + (1-n) \sigma_{\infty},
\end{equation}
where $n$ is the fraction of tiles in the annulus.

Now it is shown in appendix~\ref{fixe}, equation~(\ref{frac:tiles}),
that the fraction of tiles in the region of effective codimension
$c_{eff} = x D$ is $n(x)=\psi(x)/D$ where $\psi$ is a regular function
tending to $2/3$ when $x$ goes to 0. Hence the fraction of tiles 
in the annulus
\begin{equation}
n \simeq \frac{2}{3} \frac{c_0}{D}.
\end{equation}
Finally, we estimate
\begin{equation}
\bar{\sigma}_D \simeq \sinf - \frac{B}{D},
\label{asymp}
\end{equation}
where $B>0$ is a constant related to the entropy difference
($\sigma_{\infty} - \sigma_{an}$) and the chosen value $c_0$, both of
which are independent of $D$.


This result holds only for a large tiling side length $p$ because the
annulus width must be large as compared to a tile edge in order to
define properly an entropy per tile. Appendix~\ref{scaling} says
that the effective codimension $c_{eff}=c_0$ at $R_{an}$ if
$R_{an}=(1-\epsilon)R$ where 
\begin{equation}
\epsilon = \frac{\pi^2}{24} \frac{c_0^2}{(D-2)^2}
\end{equation}
The above requirement reads $\epsilon R \gg 1$, that is to say $p \gg D$
with $R=pD/\pi$.  The large $p$ limit must be taken {\em before} the
large $D$ one.  Numerical finite $D$ corrections in small $p$ tilings
turn out to be of order $1/D$ but {\em positive} (see next section).

In conclusion, the local structure, and therefore the entropy per
tile, of large $D$ tilings is independent of their size, shape,
tile fractions and boundary conditions. In all these cases, the local
structure of these tilings is similar to free boundary large $D$
tilings, where one encounters only one de Bruijn line per family in
large tiling patches~\cite{paperI}. In this sense, it is a true
thermodynamic limit.  These results will be confirmed by numerical 
simulations in~\ref{thermo_revisited}.

Henceforth, we concentrate on the study of tilings inscribed in
$2D$-gons of sides $k_i=p=1$ as representative of the whole class of
high symmetry tilings.  In particular, we focus on $p=1$ tilings to
extract (numerically) the thermodynamic limit of the entropy density.

\section{Simulations}
\label{simul}
\setcounter{equation}{0}

Monte Carlo numerical simulations~\cite{Newman} are widespread in the
fields of random tilings and quasicrystals where there exist many
difficult unsolved theoretical questions~\cite{mc}. To check our
claims of a thermodynamic limit, to obtain $B_D$ for $D > 10$, and to
get more precise numerical values than those obtained {\em via}
mean-field arguments~\cite{paperI}, we perform Monte Carlo
calculations on large $D$ tilings.  The $D \ra 2$ configuration space
is sampled for large $D$ systems {\em via} single-vertex flip
dynamics. This method is validated by the connectivity of the space of
configurations in two dimensions~\cite{Kenyon93,Elnitsky97,octo01}.

These simulations utilize fixed boundary tilings for the sake of
technical convenience. It is easier to code a fixed boundary tiling in
the memory of a computer, and the entropy is easier to estimate in
this case.  We have established that the central region of a large $D$
tiling is close to a free boundary tiling. All the statistics related
to free boundary tilings will be collected in such central
regions. Note that, as far as unstrained tilings are concerned, we
restrict our study to polygonal boundaries of side $p=1$ since we
establish that such tilings behave like large $p$ ones when $D$ goes
to infinity.

\subsection{Monte Carlo algorithm}
\label{MCalgo}

The algorithm is implemented as follows: at each Monte Carlo step,
select a vertex at random, with uniform probability; if this
vertex is flippable, then flip it; repeat sufficiently many times. The
method is validated by the ergodicity of the space of configurations
{\em via} flip dynamics: any configuration is reachable from any
initial configuration.

A key point is to check that this algorithm samples
configurations with uniform probability, since we are only
interested in configurational entropy in which all configurations
play the same role. Thus we need to establish
that this algorithm defines a Markovian process which 
satisfies the {\em detailed balance condition}~\cite{Newman}:
\begin{equation}
w(\CC_1 \ra \CC_2) = w(\CC_2 \ra \CC_1),
\label{det:bal}
\end{equation}
where $w(\CC_1 \ra \CC_2)$ denotes the transition probability from
configuration $\CC_1$ to configuration $\CC_2$. When two
configurations differ by a single flip, there is only one way to go
from one of them to the other one by a single flip. Therefore
$w=1/N_V$ where $N_V$ is the number of vertices. Since $N_V$ is
independent of the tiling, relation~(\ref{det:bal}) is established.

Note that the flip acceptance rate equals the fraction (about 31\% as
reported in section~\ref{vertex:stats}) of vertices with threefold
coordination. Vertices with more than threefold coordination cannot be
flipped by an ``elementary'' phason flip. An alternative Monte Carlo
dynamics, in which a threefold vertex is flipped in every step, does
not obey detailed balance because the number of flippable vertices is
not conserved, and hence eq.~(\ref{det:bal}) is violated.

Another point to consider in Monte Carlo simulations is the
correlation time $t_0$ (in units of Monte Carlo steps per vertex)
which measures how many steps are necessary between samples to avoid
excessive sample-to-sample correlations. Even though some recent work
brought new results for small $D$ (see~\cite{Mixing} and references
therein), this question has no rigorous definitive answer for large
$D$. However, our numerical estimations of $t_0$ are all in agreement
with the conjecture $t_0 \leq N_V/2$.

\subsection{Path counting and entropy estimate}
\label{path:count}

\begin{figure}[ht]
\begin{center}
\ \psfig{figure=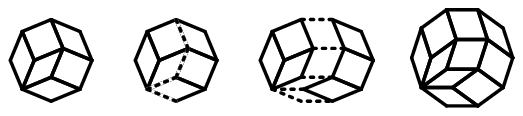} \
\end{center}
\caption{Iterative construction of fixed boundary tilings ($D \ra 2$,
  $p=1$).}
\label{iter.tilings}
\end{figure}

To estimate the entropy $\sigma_D$ of fixed boundary tilings, we shall
use a path-counting algorithm based upon de Bruijn line
enumeration~\cite{Widom97} directly derived from the general method
described in section 2.2 of paper I~\cite{paperI}.
Figure~\ref{iter.tilings} illustrates the method.  At each step, we
build a $D+1 \ra 2$ tiling from a $D \ra 2$ one.  Starting from any $D
\ra 2$ tiling, we choose a path of length $D$ along tile edges, going
from bottom to top. We cut the tiling along this path, separating the
two parts by length $1$, then draw new bonds connecting previously
identical vertices. Finally, we adjust all edge orientations to match
the set of symmetry $D+1$.

Let $\bar{P}_D$ denote the mean number of bottom-to-top paths on the $D \ra
2$ tilings:
\begin{equation}
\bar{P}_D = {1 \over B_D} \sum_{\tau=1}^{B_D} P_D(\tau),
\end{equation}
where $P_D(\tau)$ is the exact number of paths on the tiling $\tau$,
and $B_D$ is the number of such tilings. Then
\begin{equation}
B_{D+1} = \bar{P}_D B_D.
\label{iter.rel}
\end{equation}
Iterating relation~(\ref{iter.rel}), and taking the logarithm yields
\begin{equation}
\log{B_D} = \sum_{k=2}^{D-1} \log{\bar{P}_k}.
\label{logBD}
\end{equation}
For large $D$, the ratio ${{1}\over{D}} \log{\bar{P}_D}$ approaches a finite
limiting value.  Taking the limit of~(\ref{logBD}) as $D \ra \infty$,
and noting eqs.~(\ref{NT}) and~(\ref{sigmaD}), we find
\begin{equation}
\label{sigmalimit}
\lim_{D \ra \infty} \bar{\sigma}_{D}
= \lim_{D \ra \infty} {{1}\over{D}} \log{\bar{P}_D}.
\end{equation}
Hence, by accumulating statistics on the number of paths allowed on
tilings, we may evaluate the entropy.

\begin{figure}[ht]
\begin{center}
\ \psfig{figure=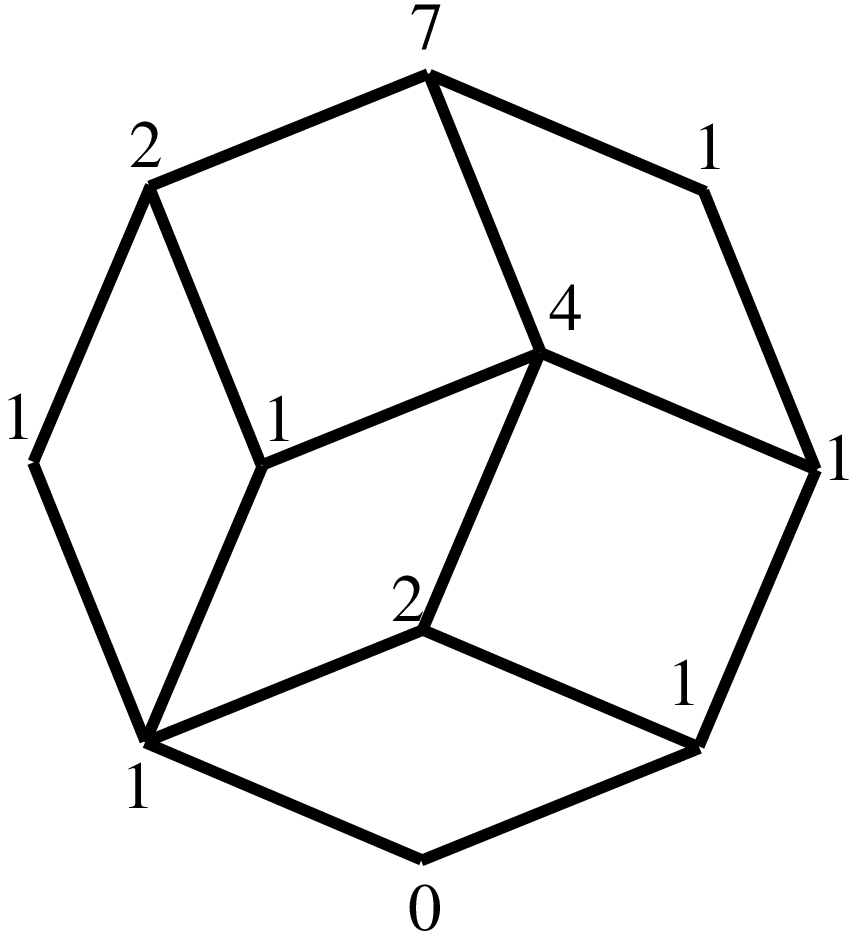,width=4cm} \
\end{center}
\caption{Path count construction, as described in the text.}
\label{pathcount}
\end{figure}

We use Monte Carlo sampling to generate an ensemble of $D \ra 2$
tilings. On each tiling $\tau$ we can quickly evaluate $P_D(\tau)$
using a generalization of the Pascal Triangle construction: starting
at the bottom of the tiling, assign each vertex an integer value equal
to the number of paths that reach it from the bottom. The value at any
vertex is iteratively the sum of the values at each prior vertex to
which it is connected (see Fig.~\ref{pathcount}). At the end of the
process, the value of the top vertex is $P_D(\tau)$.

\begin{figure}[ht]
\begin{center}
\ \psfig{figure=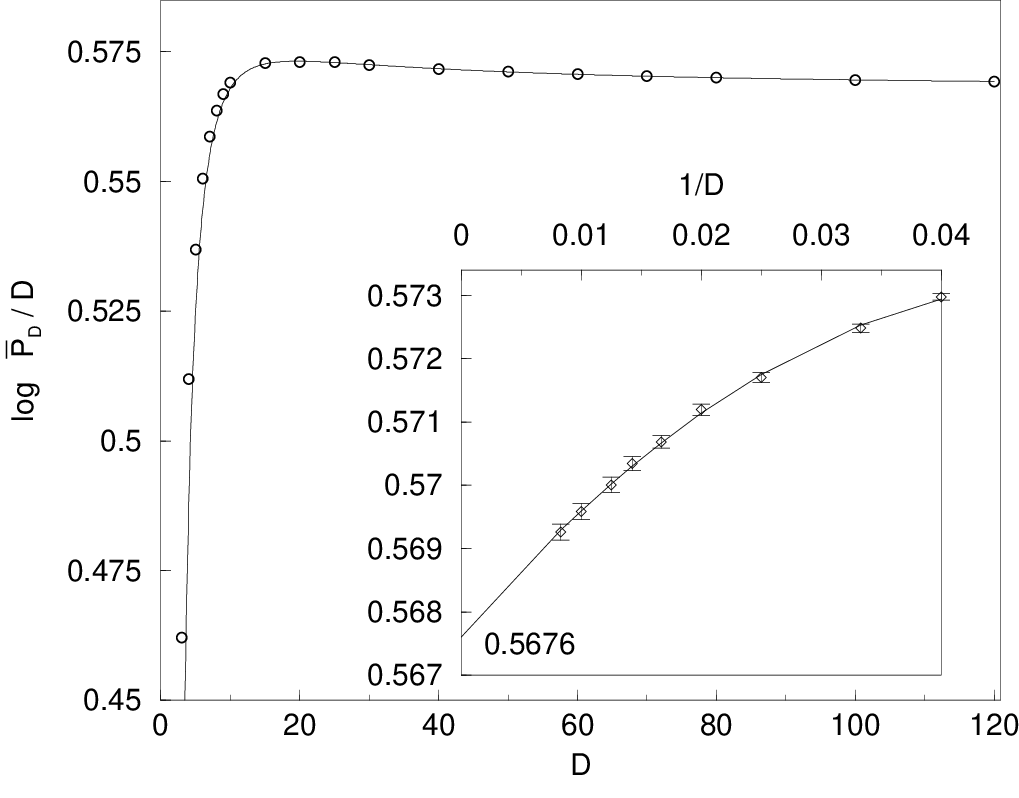,width=10cm} \
\end{center}
\caption{Value of $\displaystyle{\log \bar{P}_D /D}$ as a function of $D$ up
to $D=120$; inset: large $D$ behavior as a function of $1/D$ and
quadratic fit. The numerical values (symbols) as well as the fitted ones
(lines) present a maximum near $D=20$.}
\label{Mean}
\end{figure}

We tested this algorithm on small $D$ systems, where numerical values
are in good agreement with the exactly known mean number of paths
$\bar{P}_D=B_{D+1}/B_D$ (table~\ref{pavages.finis}). For example,
table~\ref{test.accuracy} shows the convergence towards the exact
value for $D=9$ tilings. For large $D$ systems, simulations are
our only means of obtaining information about path count statistics,
which are displayed in figure~\ref{Mean}. In this figure, error bars
are calculated as follows: the standard deviation $\Delta$ is measured
numerically. The error is estimated by $\Delta/\sqrt{I}$,
where $I$ is the number of {\em independent} measures in the
simulation (that is~\cite{Newman} its total length divided by twice
the correlation time $t_0$). Note that the actual errors calculated
in table~\ref{test.accuracy} are compatible with the previous error
bar.

This graph presents a maximum near $D=20$ which is related to finite
$D$ {\em and} finite size corrections.  Paper I~\cite{paperI} shows
that for free boundary systems, finite $D$ corrections to $\log
\bar{P}_D /D$ are of order $1/D^2$.  Section~\ref{thermo} of the
present paper argues for {\em negative} $1/D$ finite size corrections
with fixed boundary conditions in the large $p$ limit. Here, though,
we examine $p=1$ systems with one de Bruijn line per family.  Our
numerical results indicate that corrections also fall off as $1/D$ to
the first power. Combining these corrections suggests
\begin{equation}
{\log \bar{P}_D \over D} = \bar{\sigma}_{\infty} + {A \over D} - {B
\over D^2} + O({1 \over D^3}).
\label{developt}
\end{equation}

The numerical results in figure~\ref{Mean} reproduce this behavior.
All numerical data beyond $D=15$ coincide with the fitted ones up to
error bars. In particular, the maximum observed near $D=20$ is
reproduced.  We estimate the limiting value of $\displaystyle{\log
\bar{P}_D /D}= 0.5676 \pm 0.0001 $. The uncertainty estimate on this
last limit comes from excluding the data with $D=120$.
These observations provide the
first order correction to the entropy at $p=1$:
relation~(\ref{logBD}) reads
\begin{equation}
\log B_D(1) \simeq \sum_{k=2}^{D-1} k(\sinf + \frac{A}{k})
\simeq \frac{D(D-1)}{2} \sinf + A(D-2)
\end{equation}
and 
\begin{equation}
\bar{\sigma}_D(p=1) = \frac{2}{D(D-1)} \log B_D(1) 
\simeq \sinf + \frac{2A}{D}. 
\label{asymp2}
\end{equation}

Note that our simulations yield $A>0$ which means that these
corrections of order $1/D$ are {\em positive}. By contrast,
corrections to $\sinf$ in relation (\ref{asymp}) are also of order
$1/D$ but {\em negative}. The sign of these large $p$ corrections is
the combination of two effects: the paths visit low codimension
regions near the boundary, which should increase the entropy (for
example, a path running on a square grid of codimension 0 has an
entropy density of $\log 2 > \sinf$); the entropic repulsion between paths near
the top and bottom vertices, where they are crowded, decreases the
entropy. In the $p=1$ case, paths are not subject to entropic
repulsion.  Presumably the coefficient $A$ in eq.~(\ref{developt}) is
a function of $p$ and changes its sign when $p$ grows.

\subsection{Thermodynamic limit revisited}
\label{thermo_revisited}

Here we test the results of section~\ref{thermo} where it was
demonstrated that the local structure and the entropy per tile of
tilings becomes independent of boundary conditions when $D$ becomes
large. We compare relevant numerical quantities both in the
whole tiling and in the central region, that is supposed to be close
to a free boundary tiling. We concentrate on vertex coordination numbers
which are indicators of the local structure and on path counting 
which is related to entropy. In the last subsection, we focus on
strained fixed boundary tilings.

\subsubsection{Vertex statistics}
\label{vertex:stats}

Other pertinent information available from Monte Carlo simulations
concerns vertex statistics. They are indicators of the tiling's local
``microscopic'' structure. They also are a fundamental ingredient of
the mean-field approach of paper I~\cite{paperI}.  We focus on
quantities which can be computed in the mean-field theory, especially
coordination numbers and related quantities. We gather statistics both
in the whole tiling (excluding vertices actually on the boundaries)
and in a central region containing about 20\% of the
vertices. Presumably, the central region is only weakly influenced by
the fixed boundaries, and we take that data as representative of free
boundary tilings.

\medskip

\begin{figure}[ht]
\begin{center}
\
\psfig{figure=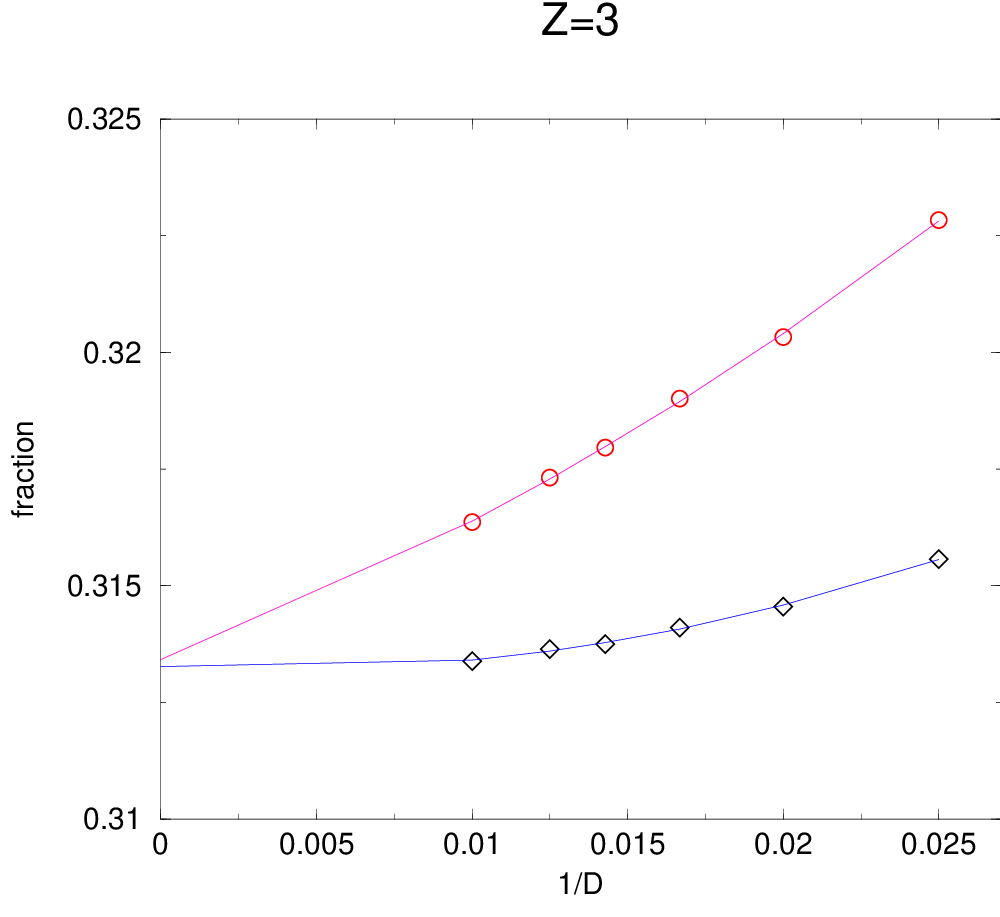,height=7.3cm}~\hspace{1cm}~\psfig{figure=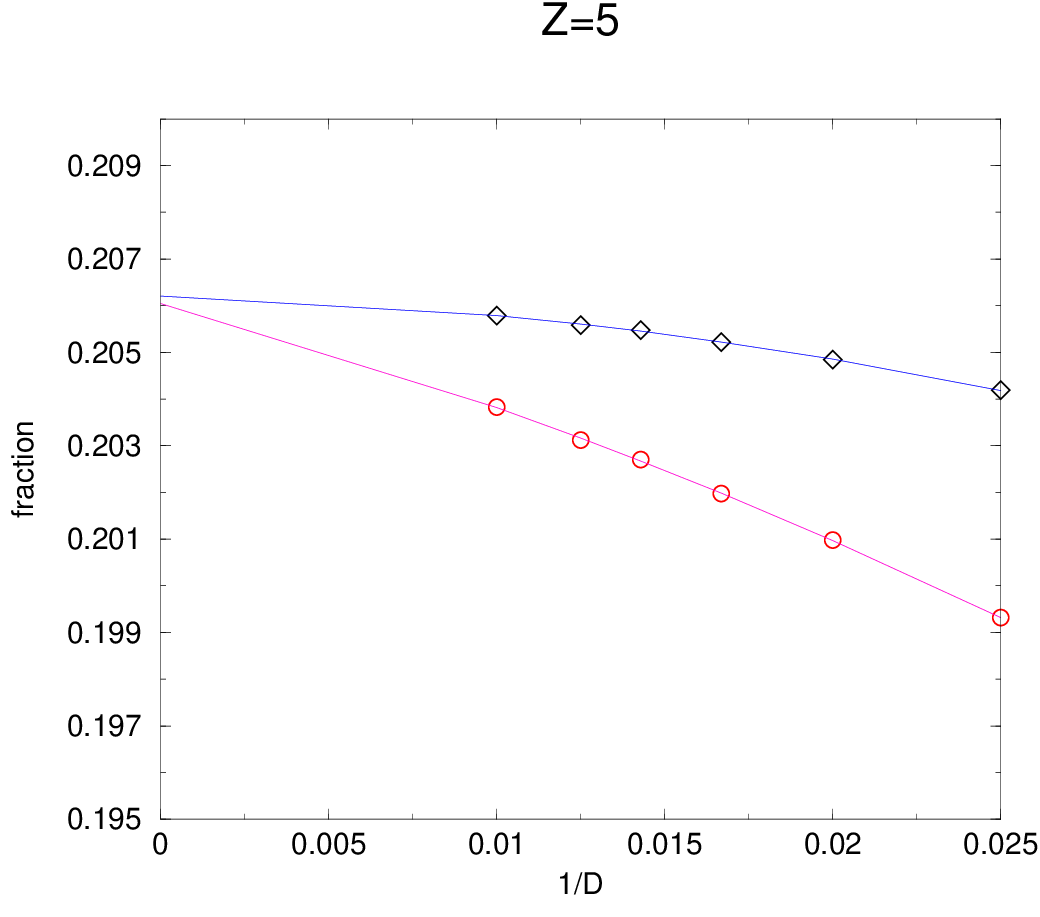,height=7.3cm}
\
\caption{Fractions of $Z=3$ and $Z=5$ vertices in $D \ra 2$ vertices
as a function of $1/D$, and quadratic fits. Diamonds concern the
central region (20\% vertices), whereas circles concern the whole
tiling. The extrapolated large $D$ values coincide.}
\label{Z_fractions}
\end{center}
\end{figure}

We have run Monte Carlo simulations up to $D=100$. After plotting the
data as functions of $1/D$, we extrapolate the limiting values {\em
via} quadratic fits.  Figure~\ref{Z_fractions} illustrates two
examples of coordination number statistics. We find the fractions of
vertices (in the whole tiling and in the central region) whose
coordination numbers $Z=3$ and the fractions whose coordination
numbers $Z=5$. For finite $D$ we see the central region has a
relatively small fraction of tiles with $Z=3$ in comparison with the
whole tiling, and a relatively large fraction with $Z=5$. The boundary
regions are thus more likely to have $Z=3$ and less likely to have
$Z=5$ than a free boundary tiling. However, the extrapolated values
for $D \ra \infty$ agree to within the accuracy of the
extrapolation. The same conclusion holds for $Z=4$ (not represented
here). For larger values of $Z$, the fractions of vertices are too
small and we did not obtain relevant measures.  Recall that the mean
value of $Z$ is exactly 4 in an infinite tiling, according to Euler's
theorem.

Hence, in the limit $D \ra \infty$ the whole tiling exhibits
coordination statistics similar to a free boundary tiling.  This point
supports the existence of a thermodynamic limit - the local structure
becomes uniform throughout the tiling, independent of proximity to the
boundaries and the local strain they create.

Other quantities we examined relate directly to mean-field theory and
were reported in paper I~\cite{paperI}. Our mean-field theory was
based upon the number of choices to be made while inscribing directed
paths on tilings. The number of choices $N_c(v)$ at a vertex $v$ is
the number $p$ of ``arms'' emerging from a vertex. The number $q$ of
``legs'' leading into a vertex together with the number of arms obey
$p+q=Z$. Paper I reports the simulated probability distribution for
$p$ and mean value of $pq$, and shows these agree well with mean-field
theory.  Here, we display in table~\ref{Z_num} the probability distribution
for $Z$, and compare it with the predictions of mean-field theory.
The agreement is also satisfactory.

\begin{figure}[ht]
\begin{center}
\
\psfig{figure=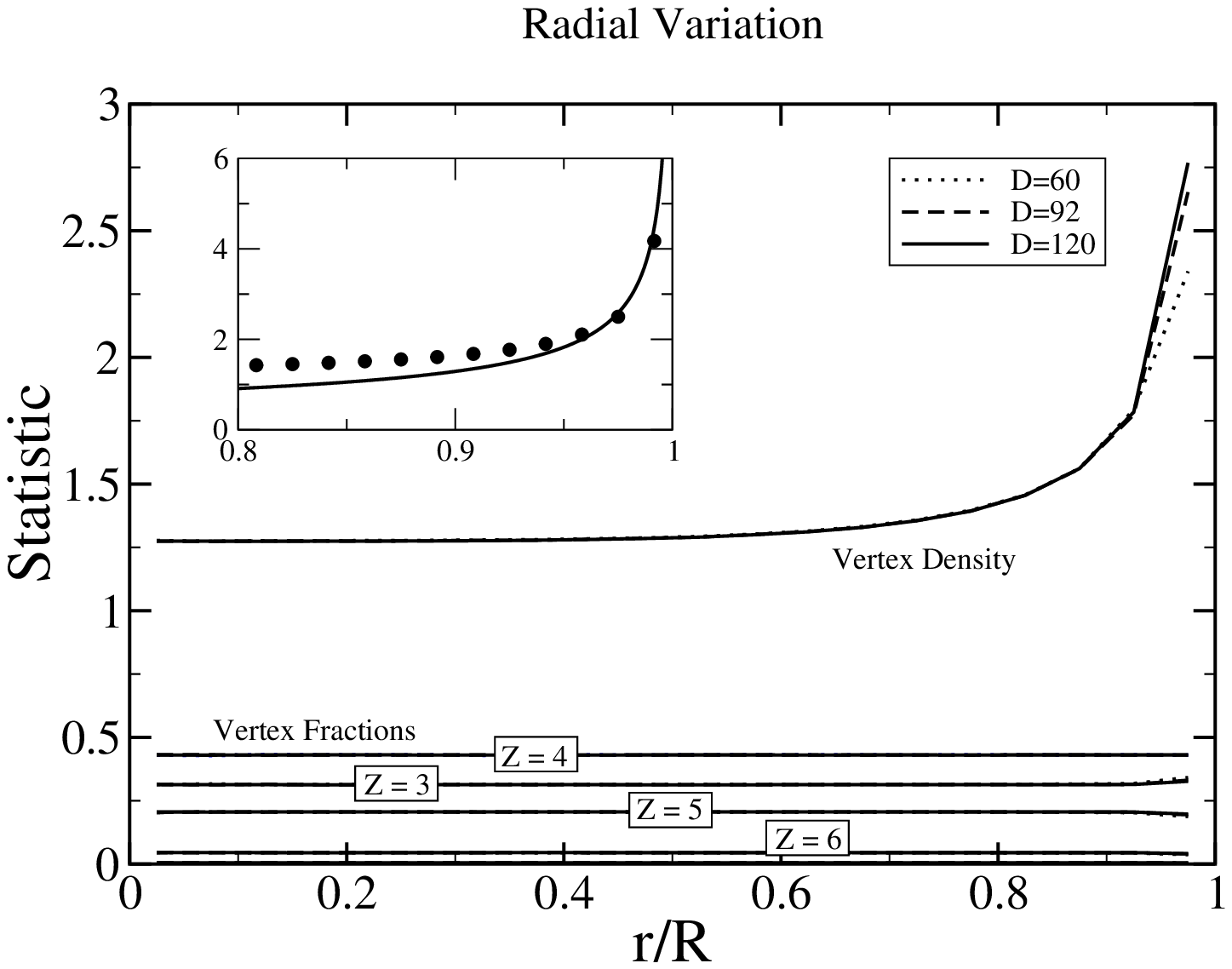,angle=-90,height=7.3cm}
\
\caption{Radial variation of vertex statistics from tiling center to edge. 
Only the vertex density shows spatial variation. Inset: vertex density
for $D$=200 (points) matches the function $1/\sqrt{6(1-r/R)}$ (line)
as $r/R\ra 1$.}
\label{rad-stats}
\end{center}
\end{figure}

An intriguing feature of the fixed boundary tilings is the divergence
of tile vertex density near the boundaries, 
\begin{equation}
d(r/R)\simeq\frac{1}{\sqrt{6(1-r/R)}},
\label{diverge}
\end{equation}
caused by the vanishing of tile area, and established in
appendix~\ref{scaling} for a particular tiling.  This effect is
plainly visible in Fig.~\ref{pavagesD2}.  Given this spatial
nonuniformity it is natural to worry about the path counting arguments
because the choice statistics might vary among different portions of
the whole tiling.  It turns out this does not happen.
Fig.~\ref{rad-stats} illustrates this point by plotting vertex
statistics averaged over different portions of the tiling.  We divided
tilings into 20 concentric circular bins and evaluated in each bin the
average vertex density and the average fractions of vertex with each
coordination number $Z$.  The results are illustrated for the cases
$D=60, 92, 120$.  Clearly the vertex density converges to a
non-constant function which diverges at the tiling boundary according
to the prediction of eq.~(\ref{diverge}).  However, the fractions of
each vertex type rapidly converge to their large $D$ limit independent
of position within the tiling. Consequently the entropy per {\em
vertex} should be uniform, even though the entropy per {\em area}
diverges due to the diverging vertex density.

\subsubsection{Path counting in the central region}
\label{paths_central}

To confirm that the entropy per tile (or vertex) is independent of
boundary conditions, we repeat calculations like
section~\ref{path:count}, but we concentrate on the central region.
It is reasonable to suppose that statistics in the central region
should match the statistics inside free boundary tilings.

\begin{figure}[ht]
\begin{center}
\
\psfig{figure=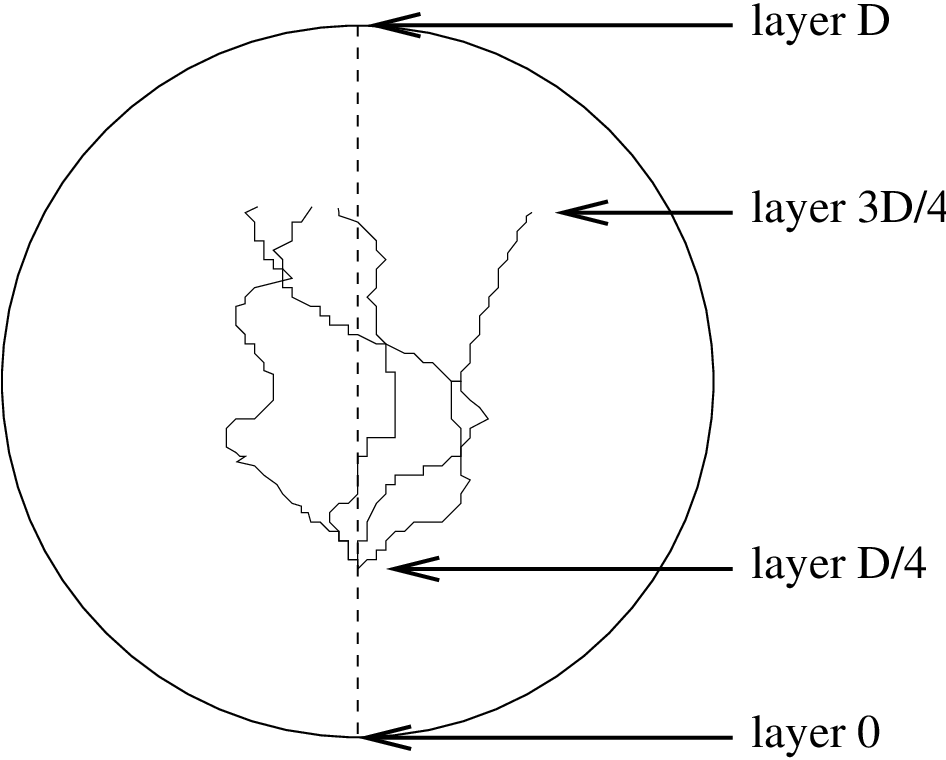,height=5cm}
\
\bigskip
\caption{Paths in the central region going from a vertex in layer
$D/4$ close to the vertical diameter to any vertex in layer
$3D/4$.}
\label{central:fig}
\end{center}
\end{figure}

On a $D\ra 2$ tiling we define $D+1$ layers of vertices as follows:
layer 0 contains only the bottommost vertex $v_0$, layer 1 contains
the vertices at distance 1 from $v_0$, layer $k$ contains the vertices
at distance $k$ from $v_0$ when one follows bottom-to-top paths.
Layer $D$ contains only the topmost vertex $v_D$.  Let $\bar{P'}_D$
denote the mean number of length $D/2$ paths from a given vertex
$v_{start}$ of layer $D/4$ to any vertex of layer $3D/4$.  For
convenience, we take $D$ a multiple of 4, and also require that
$v_{start}$ is close to the vertical diameter (see
figure~\ref{central:fig}).  Most paths stay close to this central
diameter because their typical deviation from this diameter grows like
$\sqrt{D}$ whereas the size of the tiling grows like $D$.
Consequently, $\bar{P'}_D$ counts paths in the central region.  Data
is averaged over many independent tilings.

Fitting finite $D$ data as in section~\ref{path:count}, we find the
limiting value
\begin{equation}
\lim_{D \ra \infty} \frac{\log \bar{P'}_D}{D/2} = 0.5670 \pm 0.0005.
\end{equation}
This value coincides with its fixed boundary counterpart up to error
bars.  If the path counting statistics in large $D$ fixed boundary
tilings were influenced by boundary conditions, we would have expected
the central value to differ from the value over the entire tiling.  We
conclude that the fixed boundary entropy per tile matches the free
boundary entropy in the $D\ra \infty$ limit.

\subsubsection{Phason strained tilings}
\label{phason:strain}

The data just reported in section~\ref{vertex:stats} suggests that the
tiling structure does not depend on the phason strain, since
statistics are identical in the central region and in the boundary
vicinity. We want to test this point directly by performing simulations
on phason strained fixed boundary tilings.

\begin{figure}[ht]
\begin{center}
\ \psfig{figure=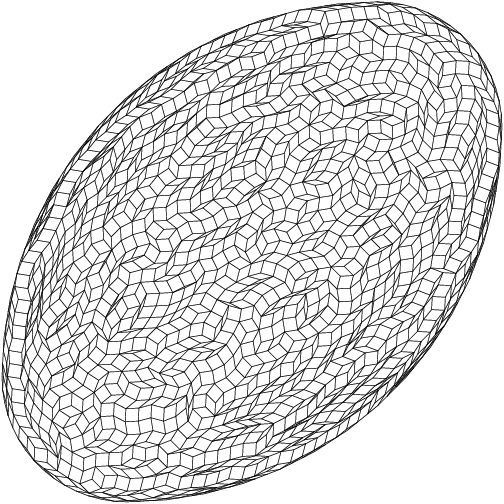,width=6.5cm}
\end{center}
\caption{Phason strained large $D$ tiling. The added paths still go from
bottom to top.}
\label{fig:strain}
\end{figure}

We consider three different forms of phason strain in our numerical
simulations: low frequency strain in which $D/2$ consecutive line
families occur with $k_i=1$ $(i=1,\ldots,D/2)$, the remaining $D/2$
consecutive line families occur with $k_i=2$ $(i=D/2+1,\ldots,D)$; the
same low frequency with a larger amplitude so that $k_i=3$
$(i=D/2+1,\ldots,D)$ replaces $k_i=2$; high frequency strain with
$k_i=1$ ($i$ odd) and $k_i=2$ ($i$ even).  Figure~\ref{fig:strain}
illustrates an equilibrated tiling with applied phason strain. The
tiling has $D=30$ and the strain is low frequency and large amplitude.

Computer simulations of path count statistics for $D=10,...,50$ reveal
no significant dependence of path count on phason strain amplitude or
frequency. The small variations seen (see table~\ref{tab:strain})
cannot be separated from finite size effects. Thus entropy does not
depend on phason strain, supporting {\em a posteriori} arguments
developed for free boundary tilings~\cite{paperI}. Recall that
strain-independence of entropy in fixed boundary tilings relies upon
strain-independence in free boundary tilings (see
section~\ref{thermo}).

\begin{figure}[ht]
\begin{center}
\
\psfig{figure=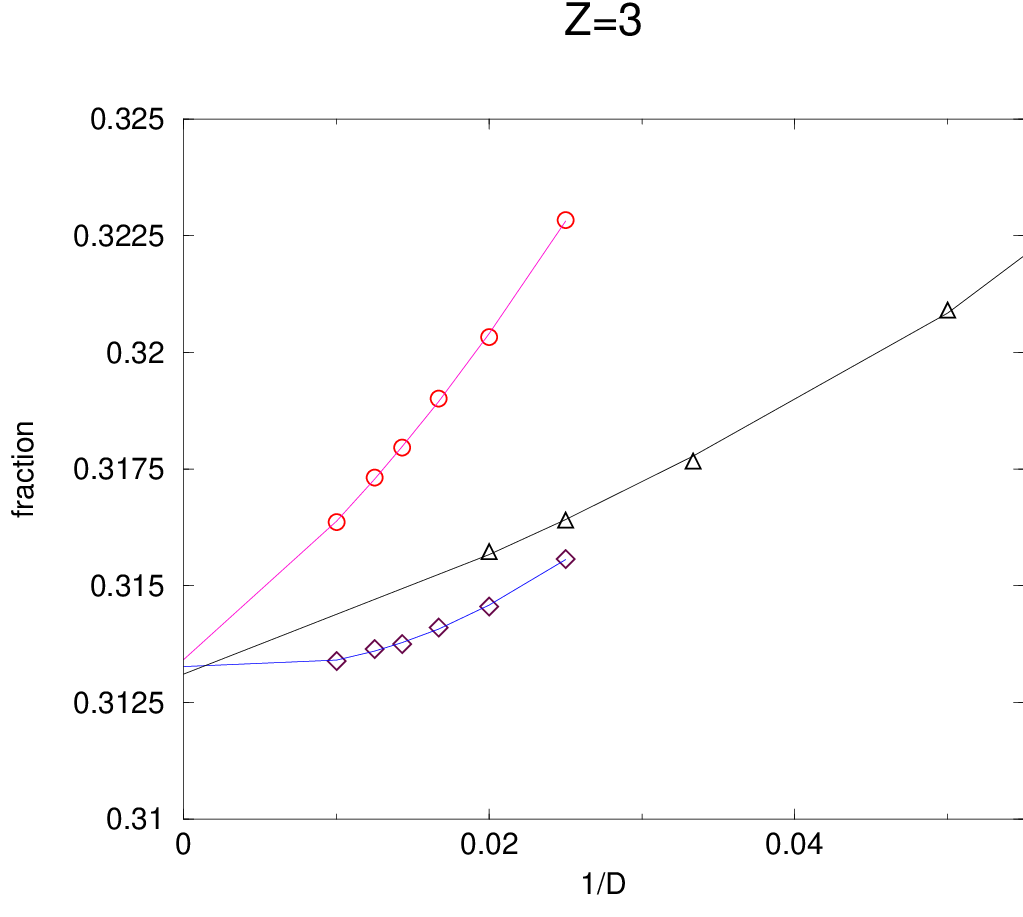,height=7.3cm}~\hspace{1cm}~\psfig{figure=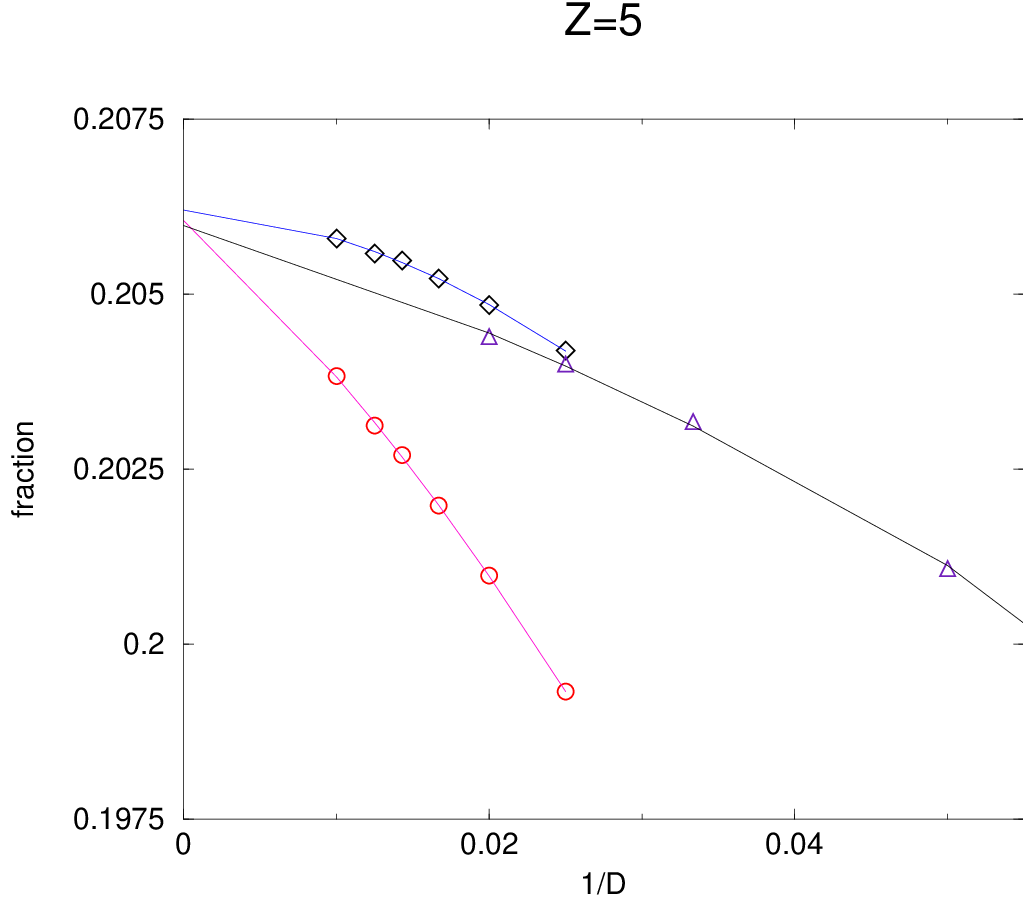,height=7.3cm}
\
\end{center}
\caption{Fractions of $Z=3$ and $Z=5$ vertices in strained tilings 
($p=1,1,\ldots,3,3,\ldots$) (triangles), together with previous
data on fixed boundary unstrained tilings (circles) and free boundary
ones (diamonds). The limiting values coincide with the
previous one, up to statistical errors. Lines are quadratic fits.}
\label{coord:strained}
\end{figure}

The same conclusion holds concerning vertex statistics on coordination
number and choice distributions. No significant differences are
found in comparison to unstrained fixed boundary tilings or free
boundary tilings (see figure~\ref{coord:strained}).  These data
support the existence of a universal local structure, independent of
strain and boundary conditions in the limit $D \ra \infty$.

\section{Conclusion}
\label{conclusion}
\setcounter{equation}{0}

This paper tackles random tilings of high symmetry with fixed boundary
conditions.  In random tiling theory, boundary conditions are crucial
because finite codimension fixed boundary tilings have a lower entropy
than free boundary ones. However, we argue that boundaries become
irrelevant when the codimension becomes large.  We also demonstrate that
$p=1$ tilings with one de Bruijn line per family have the same entropy
per tile as tilings with $p$ large and even infinite, for sufficiently
large $D$.

As a consequence, the numerical study of the entropy of large
codimension random tilings can be concentrated on fixed boundary
tilings filling a $2D$-gon of side lengths set to 1. We perform exact
enumeration for $D\leq 10$, thanks to an analogy between fixed
boundary tilings and some class of sorting algorithms. We use Monte
Carlo simulations for larger tilings, up to $D=120$.  In both cases,
the fact that tilings have a fixed polygonal boundary greatly
simplifies their encoding and their manipulation in the memory of a
computer. Our Monte Carlo data analysis is based on the same iterative
process (i.e. concentrating on values of $\bar{P}_D$) as our
mean-field theory in paper I~\cite{paperI}.  We obtain a very
accurate estimate of the entropy per tile, $\bar{\sigma}_{\infty} =
0.5676 \pm 0.0001 $, compared to the approximate mean-field value
$\sigma_{\infty}^{MF} \simeq 0.598$.

Insensitivity of the entropy per tile and vertex statistics to
boundary conditions at large $D$ restores a thermodynamic limit which
does not exist at finite $D$.  However, caution is required. Even if
``topological'' quantities such as entropy and vertex statistics
become homogeneous in the tiling, its ``metric'' properties remain
heterogeneous, as discussed in section~\ref{vertex:stats}.
Consequently we find the entropy per vertex is homogeneous while the
entropy density per area diverges near the tiling boundary.

Many of our methods and results can be generalized to higher
dimensional tilings. However, the generalization may not be simple to
implement in practice.  Indeed the de Bruijn directed paths become
directed surfaces built of faces of rhombohedra. The description of
these surfaces in terms of successive choices must be generalized,
which highly complicates their enumeration.

The restoration of the thermodynamic limit in the high symmetry limit
should remain valid in three- or higher-dimensions, because the vision
of large regions with at most one de Bruijn {\em surface} per family
still holds. The explicit proof should be similar to the
appendix~\ref{fixe}, with only minor changes to account for the
possible spontaneous decomposition into regions of lower effective
codimension (as in the ``arctic octahedron'', see Ref.~\cite{4to3}).
Presumably, the effective codimension still goes to infinity for
almost all tiles.

One of the main motivations for the study of high codimension random
tilings was the hope that mean-field theory might become exact in the
high codimension limit. The principal conclusion of this numerical
study, together with the mean-field results of paper I, is that the
mean-field theory does {\em not} become exact, at least in the
simplest version which neglects vertex correlations. Rather, the high
codimension limit is quite nontrivial, and an exact analytic solution
is yet to be found.

\section*{Acknowledgments}
We thank Chris Henley and Pavel Kalugin for useful discussions.  This
research is supported in part by the National Science Foundation under
grants DMR-0111198 and INT-9603372 and by the CNRS.

\appendix

\section{Thermodynamic limit of two-dimensional fixed boundary
  tilings}
\label{fixe}

In this technical appendix, we demonstrate that the fixed boundary
entropy $\bar{\sigma}_D$ attains a finite limit $\sinfbar$ when $D
\ra \infty$, and that this limit coincides with the free boundary entropy
$\sinf$. Moreover, we argue that this limit is shared with finite side
polygonal boundaries as well as non-diagonal tilings.

Our demonstration relies on the variational principle introduced in
references~[\cite{matheux,Bibi97B}] to characterize the entropy
gradient between the center and the boundary in typical tilings with
given fixed boundaries. Typical tilings are those which maximize an
entropy functional defined as the integral over the tiling of a local
entropy per tile; This local entropy is the {\em free}-boundary
entropy calculated with the local fractions of tiles. The entropy per
tile of fixed boundary tilings is the maximum of this functional.

We show below that, given any domain in the tiling, when $D$ goes to
infinity, the tiling in this domain is a piece of (free) $D' \ra 2$
tiling, such that $D'$ goes to infinity as $D$ does. Hence the
local entropy per tile tends to the large $D$ free boundary
entropy $\sigma_{\infty}$ (nearly) everywhere and fixed boundary
tilings have the same entropy as free boundary ones.

We first consider the case where the sides of the polygonal boundary
share the same length $p$ (diagonal tilings), and take the large $p$
limit before taking the large $D$ limit. Secondly, we consider the
case were $D$ becomes large at fixed $p$. We also discuss the
non-diagonal case where these side lengths might be different. Our
presentation is heuristic but can be made rigorous. Finally, we
examine scaling laws for several quantities studied in the paper,
such as the effective codimension, when they are written as functions
of the radius $r$ from the tiling center.

\subsection{Diagonal tilings}
\label{diagtilings}

\begin{figure}[ht]
\begin{center}
\ \psfig{figure=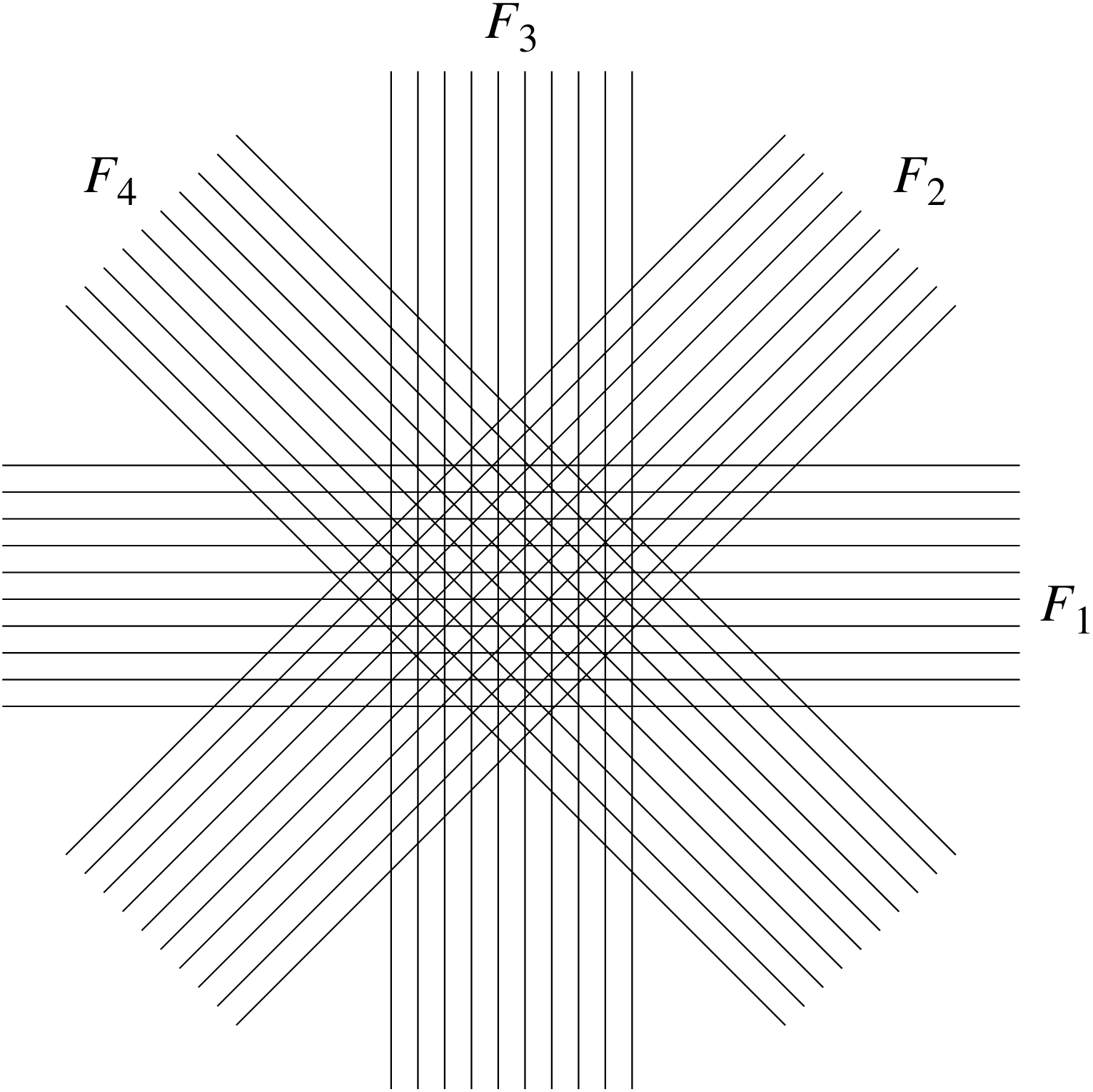,height=6cm} \hspace{1.5cm} \psfig{figure=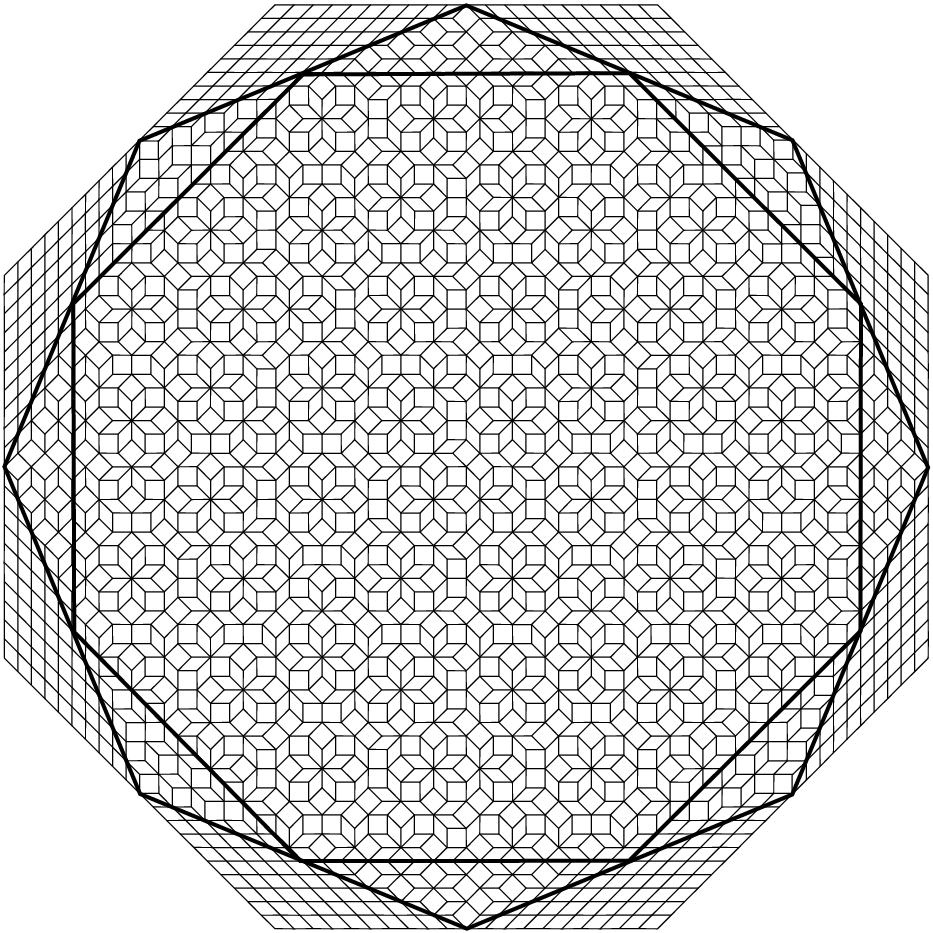,height=6cm} \
\end{center}
\caption{The de Bruijn straight lines (left) and the
corresponding rhombus tiling in the $4 \ra 2$ case with $p=20$ (for
clarity's sake, we have only drawn 10 lines per family instead of 20
on the left figure). The regions where only 3 or even 2 families
intersect clearly appear on the left figure. The corresponding regions
are delimited by the two internal octagons on the right figure: in the
outer region ${\cal A}_2$, only 2 families meet; In the intermediate
one ${\cal A}_3$, 3 families meet; And in the central region ${\cal
A}_4$, all 4 families intersect.
\label{DeBruijn4.2}}
\end{figure}

Our demonstration uses a particular tiling denoted as $\TC_0$.  It is
the dual of a particular de Bruijn grid~\cite{paperI,debruijn},
consisting of $D$ families of $p$ {\em straight} lines equally spaced
out and forming a regular fan~\cite{fan} (see
figure~\ref{DeBruijn4.2}, left). The de Bruijn families are denoted by
$F_1,F_2,\ldots,F_{D}$, and are labeled counterclockwise so $F_k$
makes an angle $k \pi / D$ with an arbitrary reference
direction. Figure~\ref{DeBruijn4.2} displays such a grid and its dual
tiling. For finite $D$, such a tiling is known~\cite{Bibi97B} not to
maximize the entropy functional among the tilings with the same boundary
(see below).  By construction, the tiling $\TC_0$ is divided
into domains where only a fraction of the de Bruijn families
intersect.  In each domain, the tiling is homogeneous, with a well
defined local entropy per tile at large $p$. Since only $D'<D$ de
Bruijn families intersect in such a domain, this entropy {\em a
priori} differs from the free boundary $D \ra 2$ one.  However, by
rotational symmetry, all the domains where exactly $D'$ families
intersect have the same local entropy. In the following, the union of such
domains where only $D'$ families meet will be denoted by ${\cal
A}_{D'}$ and called the region of ``effective dimensionality'' $D'$.

Define $S_0^{D \ra 2}$ as the entropy functional evaluated at $\TC_0$.
It satisfies $S_0^{D \ra 2} \leq \bar{\sigma}_{D}$.  Then
\begin{equation}
\lim_{D \ra \infty} S_0^{D \ra 2} \leq \bar{\sigma}_{\infty}
\leq \sigma_{\infty}.
\label{1st.ineq}
\end{equation}
Next we intend to show that for any real number $h < 1$,
\begin{equation}
\lim_{D \ra \infty} S_0^{D \ra 2} \geq h \; \sigma_{\infty}.
\label{alpha.ineq}
\end{equation}
It then follows that 
\begin{equation}
\lim_{D \ra \infty} S_0^{D \ra 2} \geq \sigma_{\infty},
\end{equation}
and, owing to relation~(\ref{1st.ineq}), that 
\begin{equation}
\sigma_{\infty} \leq 
\lim_{D \ra \infty} S_0^{D \ra 2} \leq \bar{\sigma}_{\infty}
\leq \sigma_{\infty},
\end{equation}
and therefore that the 3 involved quantities are equal.

Let us prove the above statement~(\ref{alpha.ineq}). We use the fact
(see section~\ref{phason:strain} and also paper I\cite{paperI}), that
the entropy $\sigma_{\infty}$ of free boundary tilings does not depend
on strain in the large $D$ limit. We show that the local entropy of
$\TC_0$ equals $\sigma_{\infty}$ in (nearly) all regions $\AC_{D'}$.

\begin{figure}[ht]
\begin{center}
\ \psfig{figure=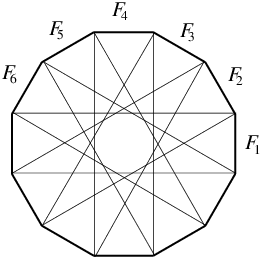,height=4cm} \hspace{1cm} 
\psfig{figure=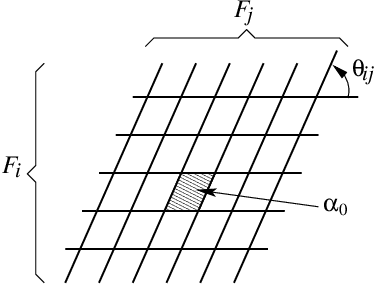,height=4cm} \
\end{center}
\caption{Left: Schematical representations of the $D$ families of de
Bruijn lines and of their intersections, to be compared to
fig.~(\ref{DeBruijn4.2}), left.  When $D$ tends to infinity, the
polygonal boundary tends toward a circle of circumference $2 D p$ and
the polygonal central region ${\cal A}_D$ tends toward a circle a diameter
$p$. Right: Two families of de Bruijn lines making an angle
$\theta_{ij}$.  The number of intersections of these two families in a
region equals its area $\alpha$ divided by the area $\alpha_0$ of
the grayed unit cell.
\label{2familles}}
\end{figure}

First we estimate how many tiles each region contains.
Since a tile is defined as the intersection of two de Bruijn lines, we
will count the number of such intersections in a given region, by
calculating the number of intersections per unit area in each region
on the one hand and the area of the regions on the other hand.
However we work not in the ``tiling metric'' but in the ``grid
metric'' instead. That is to say, in the figure where the de Bruijn
lines were originally drawn straight (left-hand representation in
fig.~\ref{DeBruijn4.2}).
Figure~\ref{2familles} (left) displays a schematic representation of a
grid were the different regions under consideration clearly appear.

The number of intersections in a region $\AC$ of area $\alpha$ is the
sum over all pairs of de Bruijn families of the number of
intersections of a pair. Consider a given pair of families $F_i$ and
$F_j$, $i<j$, which make an angle $\theta_{ij}$. The number of such
intersections is the area $\alpha$ divided by the area $\alpha_0$ of
the unit cell of the lattice defined by these (only) two families
(figure \ref{2familles}, right). If the distance which separates two
lines of a family is set to 1, then $\alpha_0 = 1/|\sin
\theta_{ij}|$. Hence the total number of intersections is
\begin{equation}
\sum_{i<j} {\alpha \over 1/|\sin \theta_{ij}|} = \alpha \sum_{i<j} |\sin
\theta_{ij}|,
\end{equation}
where the indices $i$ and $j$ run over the families present in $\AC$.

As displayed in figure~\ref{2familles} (left) a region ${\cal A}_{D'}$
is a crown made of $2D$ kites (or triangles for the inner crown) of
equal areas. A kite (or triangle) of ${\cal A}_{D'}$ will be denoted
by $\KC_{D'}$ and its area by $\kappa_{D'}$. Henceforth the area of
${\cal A}_{D'}$ is equal to $\alpha_{D'} = 2D \kappa_{D'}$. In
$\KC_{D'}$, there are $D'$ {\em adjacent} families of lines, for
example $F_1,\ldots,F_{D'}$. The angles that they make relatively to
an arbitrary reference direction are therefore $\theta_l = \theta_0 +
{l \pi / D}$, $l=1,\ldots,D'$ and the number of intersections in
$\KC_{D'}$ is
\begin{equation}
Q_{D'} = \kappa_{D'} \left( \sum_{1 \leq i<j \leq D'} \sin {(j-i) \pi
\over D} \right) .
\label{Q}
\end{equation}
When $D$ goes to infinity, $x=D'/D$ becomes a continuous variable
which represents a ``fraction'' of the $D$ families of lines. For
example, the previous sum in eq.~(\ref{Q}) can be estimated by an
integral on the variables $y=j/D$ and $z=i/D$. If $E(x)$ denotes the
integral part of $x$, then
\begin{equation}
\phi_D(x) \equiv
\sum_{1 \leq i<j \leq E(x D)}  \sin {(j-i) \pi \over D} 
\simeq D^2 \int_0^{x} dy \int_0^y dz \sin \pi (y-z)  \\
= {D^2 \over \pi^2 } \left[ \pi x -\sin \pi x \right].
\end{equation}			

The central region ${\cal A}_D$ plays a particular role
among the regions ${\cal A}_{D'}$: when $D$ is large, the fraction of
tiles in ${\cal A}_D$ tends to $1/2$. Indeed, as illustrated in figures
\ref{DeBruijn4.2} (left) and \ref{2familles} (left), the
central region tends toward a circle of diameter $p$. Its area is $\pi
p^2/4$, and, owing to the above result applied to the case $x=1$, the
number of intersections that it contains grows like
\begin{equation}
{D^2 \over \pi} {\pi p^2 \over 4} = {D^2 p^2 \over 4}.
\end{equation}
Since the total number of tiles grows like $\displaystyle{D^2 p^2
\over 2}$, this central region contains one half of the total number
of tiles.

By comparison, the situation is completely different in the other regions
$A_{D'}$, with $D'<D$. In this case, when $D$ goes to infinity, the
fraction of tiles in such a region vanishes. Let us first
explicitly compute the asymptotic behavior of this fraction and then
discuss how to handle this vanishing character.

\begin{figure}[ht]
\deuxfigs{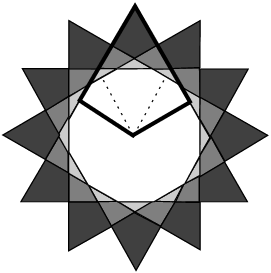}{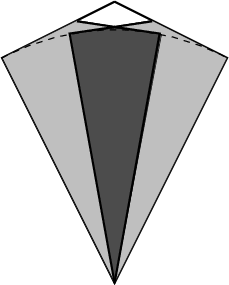}
\caption{(left) around the central region ${\cal A}_D$, there is a
first crown of triangles $\KC_{D-1}$ (light gray), a second crown of
kites $\KC_{D-2}$ (medium gray), a third crown of kites $\KC_{D-3}$
(dark gray) and so forth; in this figure, $D=6$. Right: the three
kinds of kites which define a kite-shaped region $\KC_{D'}$
(white): a big kite {\em minus} two small kites (light gray) {\em
plus} a smaller kite (dark gray).
\label{regions}}
\end{figure}

Figure~\ref{regions} (left) shows the geometry of the regions under
consideration when $D$ is finite. Around the central region ${\cal
A}_D$, there is a first crown triangular domains, then a second crown
of kite-shaped regions, and as $k$ increases, the regions ${\cal
A}_{D'}$ lie in concentric crowns around the central one. Each crown
contains $2D$ kite-shaped (or triangle-shaped) regions. As
figure~\ref{regions} (right) illustrates, such a kite can be seen
as a greater rectangular kite\cite{rectkite} {\em minus} two equal
rectangular kites. These two latter kites have a non empty
intersection, which is itself an even smaller kite.  If the areas of
these kites are respectively denoted by $\delta_{D'}$, $\delta_{D'+1}$
and $\delta_{D'+2}$, then the area of ${\cal K}_{D'}$ is
\begin{equation}
\kappa_{D'}=\delta_{D'} - 2 \delta_{D'+1} + \delta_{D'+2}.
\end{equation}

Now, for each of these 4 kites, its edges are two rays of the
circular central region and two tangents to this central region,
perpendicular to these two latter rays. Since the central region has diameter
$p$, if the two rays make an angle $\theta$, the area of the kite is
\begin{equation}
\delta(\theta)={p^2 \over 4} \tan\left( {\theta \over 2} \right).
\end{equation}
For the domain ${\cal K}_{D'}$, $\theta=\pi (D-D')/ D$ for the
greater kite.  Henceforth,
\begin{equation}
\delta_{D'}={ p^2 \over 4} \tan \left( \pi {D-D'\over 2 D} \right).
\end{equation}
The same formula holds for $\delta_{D'+1}$ and $\delta_{D'+2}$. Thus
if we set $k=D-D'$,
\begin{equation}
\kappa_{D'}=
{p^2 \over 4} \left[
\tan \left( {k \pi \over 2 D} \right)
-2 \tan \left( {(k-1) \pi \over 2 D} \right)
+\tan \left( {(k-2) \pi \over 2 D} \right) \right].
\end{equation}
Fixing the ratio $x=D'/D$, and approximating the above as a second
derivative, we finally get 
\begin{equation}
\kappa_{D'} = \kappa_{E(xD)} \simeq \pi^2 {p^2 \over 8} {1 \over D^2} 
\tan \left( \pi (1-x)/ 2  \right) \left[ 1 +
\tan^2 \left( \pi (1-x) / 2 \right) \right].
\label{alpha.k} 
\end{equation}
Thus the number of tiles in $\AC_{E(xD)}$ is
\begin{equation}
N_{E(xD)} = \alpha_{E(xD)} \phi_D(x) = 2D \kappa_{E(xD)} \phi_D(x).
\label{n:tiles}
\end{equation}
For later reference we also calculate the average area of a tile in
$\AC_{E(xD)}$,
\begin{equation}
\sigma(x)=\frac{\sum_{1\leq i,j \leq E(xD)} \sin^2\frac{(j-i)\pi}{D}}
{\sum_{1\leq i,j \leq E(xD)} \sin\frac{(j-i)\pi}{D}}
\simeq \frac{\cos(2\pi x) - 1 + 2 \pi^2 x^2}{8(\pi x - \sin \pi x)}.
\label{a:tiles}
\end{equation}
Finally, the fraction of tiles in the region
${\cal A}_{E(xD)}$ is
\begin{equation}
\label{frac:tiles}
n_{E(xD)} = \frac{N_{E(xD)}}{N_T}= \frac{N_{E(xD)}}{D^2 p^2 /2} \simeq
{1 \over 2D } \tan \left( \pi (1-x)/ 2  \right) \left[ 1 +
\tan^2 \left( \pi (1-x) / 2 \right) \right]
\left[\pi x - \sin \pi x
\right]
\end{equation}
when $D$ goes to infinity, since the number of tiles in the whole
tiling is $N_T = D^2 p^2 /2$.

As expected, these fractions vanish but
they can be added to get a non vanishing number of intersections. More
precisely, let us fix a real number $0 \leq g \leq 1 $ and let us
compute the fraction of tiles which lie in regions where at least
$E(g D)$ families of lines intersect:
\begin{equation}
{1 \over 2} + \sum_{l=E(g D)}^{D-1} n_{l} \simeq 
{1 \over 2} + {1 \over 2} \int_{0}^{1-g} dx 
\tan \left( \pi x / 2 \right)
\left[ 1 + \tan^2 \left( \pi x / 2 \right) \right] \left[ \pi (1-x) - 
\sin \pi x \right] 
\equiv {1 \over 2} + {1 \over 2} f(1-g).
\end{equation}
The above integral is a continuous function $f(g)$ for $g \in [0,1]$.
It fulfills the
required condition $f(0)=1$ (which means that the whole tiling
contains a fraction 1 of tiles!).  Moreover, $f(g) < 1$ when $g >
0$. Thus for any $h < 1$, there exists a real number $g > 0$ such that
a fraction $h$ of the tiles lie in regions where at least $E(g D)$
families of lines meet.  Since $g > 0$, when $D$ goes to infinity,
the number of families of lines also goes to infinity in such
regions. 

Moreover, appendix~\ref{encadre} demonstrates
that the local tilings in such regions are true large codimension
ones. Hence everywhere in the above region, the local entropy tends to
the free boundary entropy $\sigma_{\infty}$. A fraction $h$
of the tiles lie in regions where the local entropy per tile tends to
$\sigma_{\infty}$ when $D \ra \infty$.  We get~\cite{Bibi97B} the
expected relation~(\ref{alpha.ineq}) and we conclude that free and
fixed boundary entropies coincide.

These arguments can also be extrapolated to finite $p$ tilings
provided $D$ is large. Indeed, if $p$ is finite, a domain $\KC_{D'}$
can be very small and can contain very few tiles (and even no tile at
all).  Nevertheless in the previous demonstration, regions
$\AC_{E(xD)}$ with $x_0 \leq x \leq x_0 + \delta x$ can be put
together into larger regions, to which all the previous arguments can
be applied.

Finally we remark that the tiling $\TC_0$ which was not assumed {\em a
priori} to maximize the entropy functional indeed does so at the large
$D$ limit. Therefore it should be close to generic and should give a
good idea of the macroscopic structure of large $D$ generic tilings,
in particular as far as the scaling laws of section~\ref{scaling}
are concerned.

\subsection{Non-diagonal tilings}
\label{nondiagtilings}

For large $p$ non-diagonal tilings with side lengths $k_i = x_i p$,
where $\alpha \leq x_i \leq \beta$, we use a particular tiling
$\TC'_0$ which is a variation of $\TC_0$. Its overall definition is
the same except that the line spacing $l_i$ depends on the de Bruijn
family $F_i$. We follow the main steps of the prior demonstration, in
particular the calculation of the areas $\kappa_{D'}$ or
$\alpha_{D'}$.

We choose $l_i = p/(x_i p -1) \simeq 1/x_i$ so that the width $l_i
(k_i-1)=p$ of the family $F_i$ is independent of $i$ in the grid
representation (figure~\ref{2familles}, left). Therefore the areas
$\alpha_{D'}$ remain unchanged as compared to the previous section.

On the other hand, the density of intersections in each domain
$\KC_{D'}$ will vary because of varying line spacing. In particular
the area $\alpha_0$ of the unit cell now depends on $i$ and $j$:
\begin{equation}
\alpha_0(i,j) = \frac{l_i l_j}{| \sin \theta_{ij} |} = \frac{1}
{x_i x_j | \sin \theta_{ij} |}.
\end{equation}
The number of intersections~(\ref{Q}) becomes
\begin{equation}
\tilde{Q}_{D'} = \kappa_{D'} \left( \sum_{1 \leq i<j \leq D'} x_i x_j
\sin {(j-i) \pi \over D} \right) \leq \beta^2 Q_{D'} .
\end{equation}
Henceforth, the new numbers of tiles in the crown $\AC_{D'}$ and in
the whole tiling satisfies $\tilde{N}_{D'} \leq \beta^2 N_{D'}$ and
$\tilde{N}_{T} \geq \alpha^2 N_{T}$. If we set again $D'=E(xD)$, the
new fraction of tiles in $\AC_{D'}$ satisfies
\begin{equation}
\tilde{n}_{E(xD)} = \frac{\tilde{N}_{E(xD)}}{\tilde{N}_{T}} \leq 
\frac{\beta^2}{\alpha^2} \; n_{E(xD)}.
\end{equation}
As a consequence, the outer crowns with small effective dimensionality
$D'$, which had a vanishingly small contribution to the total number
of tiles in the previous section, will again be negligible in the
present case.

To finish the proof, we must check that the remainder of the tiling has a
local entropy equal to $\sinf$. The proof is explicited in
appendix~\ref{encadre} in the diagonal case but can easily be
adapted to the present case, leading to
\begin{equation}
\frac{n'_{ij}}{(n')^*_{ij}} \geq \frac{\alpha^2}{\beta^2} \; \frac{6}{\pi^2},
\end{equation}
where the notations are defined in appendix~\ref{encadre}. Even though
the lower bound on tile fractions now depends on $\alpha$ and $\beta$
because we are dealing with a nondiagonal case, it remains finite.

\subsection{Scaling law for the effective dimensionality}
\label{scaling}

We derive the effective dimensionality $D'=D_{eff}(r)$ in the tiling
$\TC_0$ of side length $p$ at the large $D$ limit.  Here $0<r<R$ is the
distance from the tiling center $O$ and $R\simeq p D / \pi$ is the
radius of the tiling.  Assume that in a large $D$ tiling, a region of
effective dimensionality $D_{eff}$ is an annulus $\AC_{D_{eff}}$ of
center $O$, the tiling center, and of radii $r$ and $r+\delta r$.
Hence $D_{eff}(r+\delta r) = D_{eff}(r)-1$ (the effective
dimensionality {\em decreases} with increasing $r$), that is to say
\begin{equation}
\delta r \frac{d}{dr} D_{eff}(r) = -1.
\label{ode}
\end{equation}

Now we need to compute $\delta r$. Following appendix~\ref{diagtilings} we
write $x=D_{eff}/D$.  If we knew the area $s(x)$ of $\AC_{D_{eff}}$
{\em in the tiling metric}, then we could extract $\delta r$ from
$s(x)=2\pi r \delta r$, in order to write
\begin{equation}
s(x) \frac{d}{dr} D_{eff}(r) = -2\pi r.
\end{equation}
But we can easily extract $s(x)$ from equations~(\ref{n:tiles})
and~(\ref{a:tiles}) for the number and area of tiles in $\AC_{E(xD)}$
to find
\begin{equation}
s(x) = \frac{Dp^2}{32} \tan \left( \pi (1-x)/ 2  \right) \left[ 1 +
\tan^2 \left( \pi (1-x) / 2 \right) \right] 
\left[ \cos(2\pi x) - 1 + 2 \pi^2 x^2 \right]. 
\end{equation}
It is convenient to introduce dimensionless variables which remove the
divergences for large $D$ and $p$. Specifically, we introduce
$\hat{r}\equiv r/R$ and $\hat{s}(x)\equiv s(x)/(Dp^2)$, and we recall that
$D_{eff}(r)/D=x$ and $R=pD/\pi$ to obtain
\begin{equation}
\hat{s}(x) \frac{d}{d\hat{r}} x = -\frac{2}{\pi} \hat{r},
\label{ode-scale}
\end{equation}
which is completely equivalent to eq.~(\ref{ode}).

We can solve eq.~(\ref{ode-scale}) by direct integration to obtain the
radius $\hat{r}$ corresponding to a given effective dimension $x$,
\begin{equation}
\hat{r}^2(x)-1 = -\pi \int_0^x s(x')dx' = 
{{1}\over{16}} (-14+2\cos{(\pi x)}+\pi x \csc^2({{\pi x}\over{2}})
(\pi x + 2 \sin{\pi x}))
\label{rofx}
\end{equation}
or we can invert the solution and obtain (after returning to unscaled
variables)
\begin{equation}
D_{eff}(r)=D \gamma(r/R)
\end{equation}
with $\gamma(\hat{r})$ the inverse of $\hat{r}(x)$.  Although the
solution in eq.~(\ref{rofx}) cannot be inverted in closed form, we can
expand it for small $x$ to find values of $r$ close to the boundary
$R$. Defining the small quantity $\epsilon=1-r/R$, we have
\begin{equation}
\epsilon = {{\pi^2 x^2}\over{24}} + O(x^4),
\end{equation}
then by reversion of series obtain
\begin{equation}
\gamma(r/R) = x \approx {{\sqrt{24}}\over{\pi}} \sqrt{\epsilon}.
\end{equation}
Apparently, the effective dimensionality varies rapidly near the boundary.

We also use the notion of effective codimension
$c_{eff} = D_{eff} -2$. At large $D$, one has
\begin{equation}
\frac{c_{eff}}{D-2} \simeq \frac{D_{eff}}{D} = \gamma(r/R).
\end{equation} 
All the quantities expressed as functions of $x$ are also functions of
$r/R$. For example, the vertex density
\begin{equation}
\label{vertex:density}
d(x)=\frac{1}{\sigma(x)}\simeq \frac{2}{\pi x}=\frac{2}{\pi \gamma(r/R)}
\simeq \frac{1}{\sqrt{6(1-r/R)}}=d(r/R)
\end{equation}
at small $x$, {\em i.e.} when $r$ close to $R$.  This scaling compares
well with numerical data in section~\ref{simul}.

\section{Regularity of large codimension fixed boundary tilings}
\label{encadre}

We check here that in each region $\AC_{D'}$ of the diagonal tiling
$\TC_0$ the local entropy per tile tends to the free boundary one
$\sigma_{\infty}$ as $D\ra\infty$.  As discussed in paper
I~\cite{paperI}, a sufficient condition is that tile fractions should
be ``bounded'' on $\AC_{D'}$. This means that all local tile fractions
$n_{ij}$ have the same order of magnitude as the strain-free ones
$n_{ij}^*$. In other words, there exist a finite constant $a>0$ such
that
\begin{equation}
n_{ij} \geq a \; n_{ij}^*
\label{encadreappend}
\end{equation} 
for all $i$ and $j$. 
We want to calculate the tile fractions $n'_{ij}$ in $\AC_{D'}$ 
and compare them
to the strain-free ones. Recall that in a $D \ra 2$ tiling,
\begin{equation}
n^*_{ij} = C \; \sin\left| \pi \; \frac{j-i}{D} \right|,
\label{prop}
\end{equation}
where the constant $C$ comes from the normalization relation $\sum
n^*_{ij} =1$.

As in the previous appendix, we set $D'=E(xD)$ where
$D$ is large and $x>0$ is finite. The main difficulty in this appendix
comes from the fact that in the region $\AC_{D'}$ under consideration,
not all tile species occur. Without loss of generality, we assume that
only the tiles with indices $i,j=1,\ldots,D'$ exist. These tiles
appear in the region ${\cal A}_{E(xD)}$ with the
fractions~(\ref{prop}), but with a different normalization
constant because only some of them appear. The normalization relation now reads
\begin{equation}
\sum_{1\leq i<j \leq D'} n'_{ij}=1.  
\end{equation}
At the large $D$ limit, this sum can be replaced by an integral and we
get the tile fractions in ${\cal A}_{E(xD)}$:
\begin{equation}
n'_{ij} = \frac{1}{(D')^2} \; \frac{\pi^2 x^2}{\pi x - \sin (\pi x)} \;
\sin \left| x \pi \; \frac{j-i}{D'} \right|.
\end{equation}

Now we check that this tiling has bounded local fractions of
tiles {\em when it is considered as a $D' \ra 2$ tiling}. We need to
compare the above $n'_{ij}$ to the corresponding quantities in an
unstrained $D' \ra 2$ tilings, in other words to
\begin{equation}
(n')^*_{ij} = \frac{\pi}{(D')^2} \; \sin \left| \pi \; \frac{j-i}{D'} \right|.
\end{equation}
By convexity of the sin function on the interval $[0,\pi]$, we have: 
$\sin | x \pi (j-i)/D' | \geq x \sin | \pi (j-i)/D' |$, and
\begin{equation}
\frac{n'_{ij}}{(n')^*_{ij}} \geq \frac{x}{\pi} \; \frac{\pi^2 x^2}{\pi
x - \sin (\pi x)} \geq \frac{6}{\pi^2},
\end{equation}
which achieves the proof: $a=6/\pi^2$ in
condition~(\ref{encadreappend}). From the results of paper
I~\cite{paperI}, we conclude that in such a region $\AC_{D'}$, the
local entropy per tile equals the free boundary value $\sigma_{\infty}$
when $D \ra \infty$.

\newpage

\begin{table}[ht]
\caption{Exact (except $B_{11}/B_{10}$) data for fixed boundary
tilings, $p=1$ and $D=1, 2, \ldots, 10$.}
\vspace{0.3cm}
\begin{center}
\begin{tabular}{|l|r|r|r|r|r|}
\hline
$D$ & 1 & 2 & 3 & 4 & 5 \\
\hline
$B_D(p=1)$ & 1 & 1 & 2 & 8 & 62 \\
$\bar{\sigma}_{D}(p=1)$ & & 0 & 0.231 & 0.347 & 0.413 \\
$B_{D+1}/B_D$ & 1 & 2 & 4 & 7.75 & 14.65 \\
\hline
\hline
$D$ & 6 & 7 & 8 & 9 & 10 \\
\hline
$B_D(p=1)$ & 908 & 24,698 & 1,232,944 & 112,018,190 & 18,410,581,880 \\
$\bar{\sigma}_{D}(p=1)$ & 0.454 & 0.482 & 0.501 & 0.515 & 0.525 \\
$B_{D+1}/B_D$ & 27.20 & 49.92 & 90.85 & 164.35 & $295.97 \pm 0.04$\\
\hline
\end{tabular}
\end{center}
\label{pavages.finis}
\end{table}

\begin{table}[ht]
\caption{Exact data for fixed boundary tilings, $p=2$ and 
$D=1, 2, \ldots, 6$. }
\vspace{0.3cm}
\begin{center}
\begin{tabular}{|l|r|r|r|r|r|r|}
\hline
$D$ & 1 & 2 & 3 & 4 & 5 & 6 \\
\hline
$B_D(p=2)$ & 1 & 1 & 20 & 5,383 & 16,832,230 & 570,702,721,864  \\
$\bar{\sigma}_{D}(p=2)$ & & 0 & 0.250 & 0.358 & 0.416  & 0.451 \\
\hline
\end{tabular}
\end{center}
\label{pavages.finis2}
\end{table}

\begin{table}[ht]
\caption{Convergence of $D=9$ path count data for increasing run
length. Measured numerical errors on the latter quantity (Num. err.)
are always smaller than estimated error bars (Estim. err.).
The latter are calculated using numerically measured standard 
deviations and conjectured auto-correlation times (section~\ref{MCalgo};
[\cite{Newman}]).
}
\vspace{0.3cm}
\begin{center}
\begin{tabular}{|l|r|r|r|r|r|r|}
\hline
$N_{MC}$ & $10^3$ & $10^4$ & $10^5$ & $10^6$ & $10^7$ & exact \\
\hline
$\bar{P}_9 $ & 159.264& 164.176& 164.411 & 164.371 & 164.344 & 164.35 \\
$\log \bar{P}_9 /9$ & 0.563396 & 0.566771 & 0.566930 & 0.566903 & 0.566884 & 0.566891 \\
Num. err. & 3.5 $10^{-3}$& 1.2 $10^{-4}$& 3.9 $10^{-5}$  & 1.2 $10^{-5}$
& 6 $10^{-6}$ & \\
Estim. err. & 8.5 $10^{-3}$& 2.7 $10^{-3}$& 8.5 $10^{-4}$ & 2.7 $10^{-4}$
& 8.5 $10^{-5}$ & \\
\hline
\end{tabular}
\end{center}
\label{test.accuracy}
\end{table}

\begin{table}[ht]
\caption{The first values of the limiting distribution of coordination
numbers $Z$, obtained both in the mean-field approximation and
numerically, by Monte Carlo simulations.}
\begin{center}
\begin{tabular}{|l|r|r|r|r|r|}
\hline
$Z$ & 3 & 4 & 5 & 6 & 7 \\
\hline
Mean-field & 0.33 & 0.41 & 0.20 & 0.05 & 0.009\\
\hline 
Numerical & 0.31 & 0.43 & 0.21 & 0.04 & 0.005 \\
\hline
\end{tabular} 
\end{center}
\label{Z_num}
\end{table}

\begin{table}[ht]
\caption{Path count statistics for strained tilings.}
\label{tab:strain}
\begin{center}
\begin{tabular}{|l|r|r|r|r|}
\hline
$D$ & $p=1,1,...$ & $p=1,1,...,2,2,...$ & $p=1,1,...,3,3,...$ 
& $p=1,2,1,2,...$ \\
\hline
10  & 0.569 & 0.572 & 0.572 & 0.573 \\
20  & 0.573 & 0.572 & 0.571 & 0.574 \\
30  & 0.572 & 0.571 & 0.570 & 0.572 \\
40  & 0.572 & 0.570 & 0.570 & 0.571 \\
50  & 0.571 & 0.571 & 0.570 & 0.570 \\
\hline
\end{tabular}
\end{center}
\end{table}


\bigskip

\begin{thebibliography}{99}

\bibitem{paperI} N. Destainville, M. Widom, R. Mosseri, F. Bailly,
``Random tilings of high symmetry: I. Mean-field theory'', to be
submitted to {\em J. Stat. Phys.} (2003).

\bibitem{Elser} V. Elser, {\em Phys. Rev. Lett.} {\bf 54}, 1730 (1985).

\bibitem{debruijn} N.G. de Bruijn, {\em Ned. Akad. Wetensch. Proc.}
{\bf A84}, 39 (1981); {\em J. Phys. France} {\bf 47}, C3-9 (1986).

\bibitem{Shechtman} D. Shechtman, et al.  {\em Phys. Rev. Lett.} {\bf
53},1951 (1984).

\bibitem{Levine} D. Levine, P.J. Steinhardt, {\em Phys. Rev. Lett.}
{\bf 53}, 2477 (1984).

\bibitem{Henley91} C.L. Henley, {\em in} {\em Quasicrystals, the State
of the Art}, Ed. D.P.  Di Vincenzo, P.J.  Steinhart (World Scientific,
1991), 429.

\bibitem{elsershape} V. Elser, {\em J. Phys. A} {\bf 17}, 1509 (1984).

\bibitem{Wannier} G.H. Wannier, {\em Phys. Rev.} {\bf 79}, 357 (1950);
{\em Phys. Rev.} B {\bf 7} 5017 (E, 1973).

\bibitem{Grensing} D. Grensing, G. Grensing, {\em J. Math. Phys.}
{\bf 24}, 620 (1983); D. Grensing, I. Carlsen, H.Chr. Zapp, {\em
Phil. Mag. A} {\bf 41}, 777 (1980).

\bibitem{matheux} H. Cohn, M. Larsen, J. Propp, {\em New York J. of
Math.} {\bf 4}, 137 (1998); H. Cohn, R. Kenyon, J. Propp, {\em
J. Amer. Math. Soc.} {\bf 14}, 297 (2001).

\bibitem{Bibi97B} N. Destainville, {\em J. Phys. A: Math. Gen} {\bf
31}, 6123 (1998).

\bibitem{Widom97} M. Widom, N. Destainville, R. Mosseri, F. Bailly,
{\em in} {\em Proceedings of the 6th International Conference on
Quasicrystals} (World Scientific, Singapore, 1997).

\bibitem{Bibi00} N. Destainville, M. Widom, R. Mosseri, F. Bailly,
{\em Mat. Sci. Eng.~A} {\bf 294-296}, 409 (2000).

\bibitem{4to3} M. Widom, R. Mosseri, N. Destainville, F. Bailly,
{\em J. Stat. Phys.} {\bf 109}, 945 (2002).

\bibitem{octo01} N. Destainville, R. Mosseri, F. Bailly, 
{\em J. Stat. Phys.} {\bf 102}, 147 (2001).

\bibitem{Mosseri93B} R. Mosseri, F. Bailly, {\em
Int. J. Mod. Phys. B}, Vol 7, {\bf 6}\&{\bf 7}, 1427 (1993).

\bibitem{Bibi97} N. Destainville, R. Mosseri, F. Bailly,
{\em J. Stat. Phys.} {\bf 87}, 697 (1997).


\bibitem{Knuth92} D.M. Knuth, Axioms and Hulls, {\em in} {\em Lect. Notes 
in Computer Sci.} {\bf 606}, 35 (1992).

\bibitem{Bjorner93} {\em Oriented Matroids}, 
A. Bj\"orner, M. Las Vergnas, B. Sturmfels, N. White,
G.M. Ziegler (Cambridge University Press, 1993).

\bibitem{Kenyon93} R. Kenyon, {\em Algorithmica} {\bf 9}, 382 (1993).

\bibitem{Elnitsky97} S. Elnitsky, {\em J. Combinatorial
  Theory} A {\bf 77}, 193--221 (1997).

\bibitem{Bailey97} {\em Tilings of zonotopes: Discriminental
arrangements, oriented matroids, and enumeration}, G.D. Bailey, Ph. D. Thesis
(Univ. of Minnesota, 1997).

\bibitem{lpw} W. Li, H. Park and M. Widom,
{\em J. Stat. Phys.} {\bf 66}, 1 (1992).

\bibitem{Newman} {\em Monte Carlo methods in statistical mechanics},
M.E.J. Newman, G.T. Barkema (Clarendon Press, Oxford, 1999).

\bibitem{mc} K.J. Strandburg, L.-H. Tang and M.V. Jaric, {\em
Phys. Rev. Lett.} {\bf 63}, 314 (1989); L.J. Shaw, V. Elser,
C.L. Henley, {\em Phys. Rev. B} {\bf 43}, 3423 (1989); L.-H. Tang,
{\em Phys. Rev. Lett.} {\bf 64}, 2390 (1990); M. Oxborrow and
C.L. Henley, {\em Phys. Rev. B} {\bf 48}, 6966 (1993); F. Gahler,
Proc. ICQ5 236 (1995); D. Joseph and M. Baake, {\em J. Phys. A} {\bf
29}, 6709 (1996).

\bibitem{Mixing} M. Luby, D. Randall and A. Sinclair, {\em SIAM J. of
Comp.}  {\bf 31}, 167 (2001); N. Destainville, {\em Phys. Rev. Lett.}
{\bf 88}, 30601 (2002).

\bibitem{fan} Strictly speaking, should this fan be exactly regular,
there would be multiple intersections, for example at its very center.
To avoid this difficulty, each family must be slightly shifted by a
random distance much smaller than the interline separation. Such shifts have
been performed in the figure~\ref{DeBruijn4.2}, even though it might
not be clear because of resolution.

\bibitem{rectkite} A rectangular kite has two right angles between its
unequal sides.



\end{thebibliography}
\end{document}